\documentclass[journal,comsoc]{IEEEtran}
\usepackage{algorithm,amsbsy,amsmath,amssymb,epsfig,bbm,mathrsfs,multirow,amsthm}
\usepackage{color,amsmath}
\usepackage{algpseudocode, tabularx}
\usepackage{graphicx,caption,subcaption,array,multirow,bm}
\usepackage{xcolor, tcolorbox}
\usepackage{makecell}
\usepackage{cite}
\usepackage{threeparttable}

\usepackage[labelformat=simple]{subcaption}

\DeclareMathAlphabet{\mathpzc}{OT1}{pzc}{m}{it}

\makeatletter
\newcommand{\multiline}[1]{%
	\begin{tabularx}{\dimexpr\linewidth-\ALG@thistlm}[t]{@{}X@{}}
		#1
	\end{tabularx}
}

\makeatother
\begin{document}
\title{DDPG-E2E: A Novel Policy Gradient Approach for End-to-End Communication Systems}

\author{ Bolun Zhang, Nguyen Van Huynh, Dinh Thai Hoang, Diep N. Nguyen, and Quoc-Viet Pham
\thanks{Bolun Zhang is with the School of Engineering,  University of Warwick, Coventry CV4 7AL, UK (email: bolun.zhang@warwick.ac.uk).

Nguyen Van Huynh is with the Department of Electrical Engineering and Electronics, University of Liverpool, Liverpool L69 3GJ, UK (email: huynh.nguyen@liverpool.ac.uk).

Dinh Thai Hoang and Diep N. Nguyen are with the School of Electrical and Data Engineering, University of Technology Sydney, Sydney NSW 2007, Australia (emails: hoang.dinh@uts.edu.au; diep.nguyen@uts.edu.au).

Quoc-Viet Pham is with the School of Computer Science and Statistics, Trinity College Dublin, The University of Dublin, Dublin 2, D02 PN40, Ireland (email: viet.pham@tcd.ie). 

Preliminary results in this paper were presented at the 2023 IEEE GLOBECOM Conference~\cite{zhang2023deep}.}
}

\maketitle
\thispagestyle{empty}

\begin{abstract}
The End-to-end (E2E) learning-based approach has great potential to reshape the existing communication systems by replacing the transceivers with deep neural networks. To this end, the E2E learning approach needs to assume the availability of prior channel information to mathematically formulate a differentiable channel layer for the back-propagation (BP) of the error gradients, thereby jointly optimizing the transmitter and the receiver. However, accurate and instantaneous channel state information is hardly obtained in practical wireless communication scenarios. Moreover, the existing E2E learning-based solutions exhibit limited performance in data transmissions with large block lengths. In this article, these practical issues are addressed by our proposed deep deterministic policy gradient-based E2E communication system. In particular, the proposed solution utilizes a reward feedback mechanism to train both the transmitter and the receiver, which alleviates the information loss of error gradients during BP. In addition, a convolutional neural network-based architecture is developed to mitigate the curse of dimensionality problem when transmitting messages with large block lengths. Extensive simulations then demonstrate that our proposed solution can not only jointly train the transmitter and the receiver simultaneously without requiring prior channel knowledge but also can obtain significant performance improvement compared to state-of-the-art solutions.
\end{abstract}



\section{Introduction}
\label{Sec:intro}

\begin{figure}[!]
	\centering
	\begin{subfigure}[b]{0.45\textwidth}
		\centering
		\includegraphics[width=\textwidth]{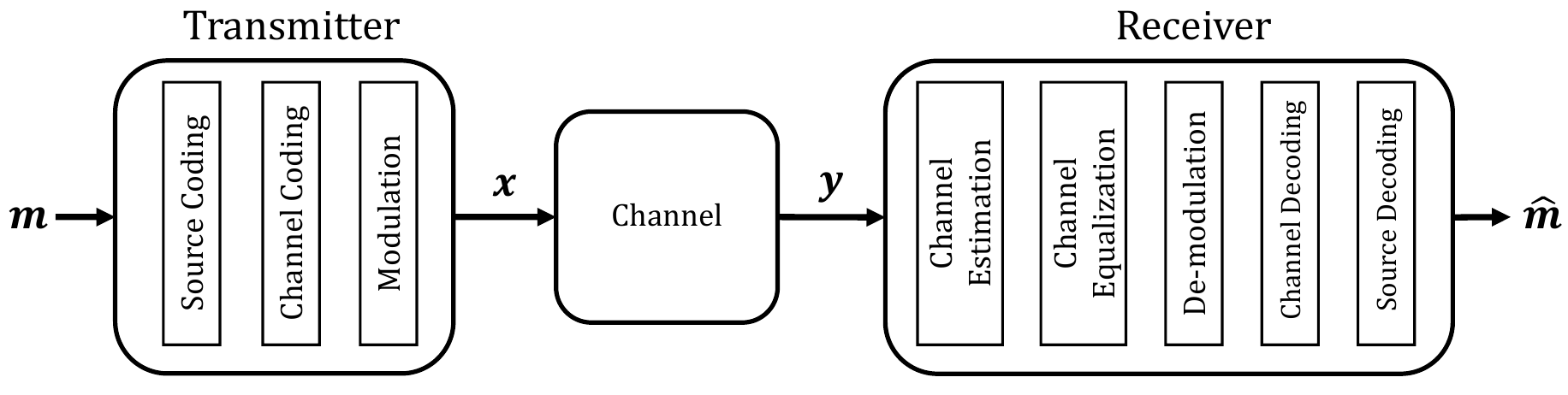}
		\caption{}
	\end{subfigure}

	\begin{subfigure}[b]{0.45\textwidth}
		\centering
		\includegraphics[width=\textwidth]{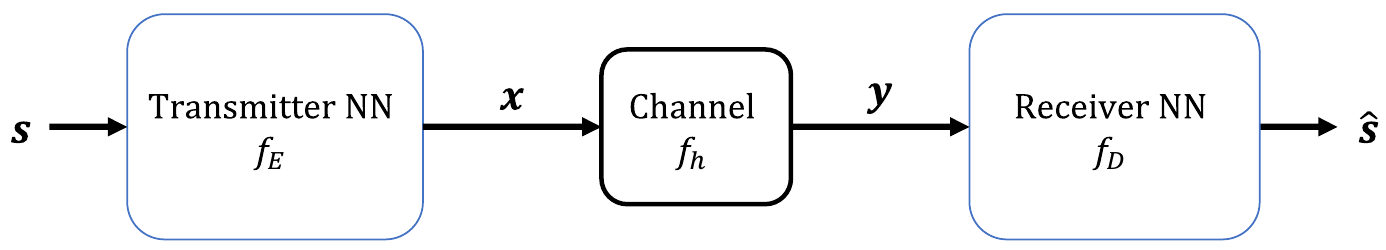}
		\caption{}
	\end{subfigure}
	\caption{Architecture of (a) traditional communication systems and (b) E2E communication systems.}
	\label{fig:architecture}
\end{figure}

\IEEEPARstart{I}{\lowercase{n}} the conventional communication system, the data transmission entails multiple signal processing blocks where each module is separately designed and optimized for a specific task, as illustrated in Fig.~\ref{fig:architecture}(a). Such a multi-block architecture has achieved reasonable performance but at the cost of increasing the design complexity. The reason is that it requires careful fine-tuning and optimizations for every individual module with different objectives. Consequently, it is difficult, if not impossible, to jointly optimize the transmitter and the receiver \cite{overAir}. In addition, these mathematical formulated modules cannot always coherently and accurately capture the actual communication conditions, thereby compromising the data transmission performance~\cite{8psk}.

Recently, deep learning (DL) technique has attracted significant attention in the field of communication systems as it provides pure data-driven solutions for handling the imperfection of communication systems. The DL-based approach interprets the end-to-end (E2E) communication system as a neural network (NN)-based auto-encoder over an interference channel \cite{intro_phy}, where both the transmitter and the receiver are represented as NNs, as depicted in Fig.~\ref{fig:architecture}(b). This E2E-based learning method theoretically assumes the availability of an explicit channel model, which is perceived as an intermediate layer connecting the transmitter and the receiver. The transmitter NN learns to map the transmitted data into an encoded vector in a smaller dimension and then sends it over the channel. The receiver NN learns to map the received signals to the estimated data. The E2E system is trained in a supervised manner to minimize its E2E loss which measures the difference between the estimated data and the transmitted data. Thus, this E2E scheme has significantly alleviated the overall design complexity by interpreting the signal processing blocks as NNs~\cite{jointtrans}. The E2E paradigm has demonstrated effective performance in simple communication scenarios, e.g., Additive White Gaussian Noise (AWGN) channel \cite{intro_phy}, while it remains challenging to apply such an E2E system to more complex wireless communication channels, such as frequency-selective channels and Rayleigh block fading channels.

The most critical shortcoming of the E2E-based learning approach is that it requires prior information of the channel between the transmitter and the receiver. Moreover, it only performs well with differentiable channel models \cite{GradientFree}. However, in practical scenarios, the communication channel is often regarded as a black box due to the non-differentiable nature of its transfer function. This characteristic poses a challenge for estimating channel gradients, especially in the context of wireless communication environments. Another critical issue with the E2E learning approach is the transmitter NN cannot be directly trained via the BP algorithm when the channel transfer function is non-differentiable. During the E2E optimization, the gradient at the receiver can be directly calculated, while the gradient at the transmitter is unavailable due to the non-differentiable channel, which hinders the calculation of the derivative of the loss function at the channel layer. Consequently, it prevents the overall realization of the E2E learning of the communication systems. In addition, the existing E2E learning scheme suffers from the curse of dimensionality. Prior works studying the E2E paradigm predominantly focus on the experiments with small block lengths (e.g., 8 information bits~\cite{intro_phy}). This occurs because the size of possible input messages grows exponentially with increasing block length when using one-hot encoding. The portion of unseen input messages would therefore exponentially increase during the training which leads to significant performance degradation. Research in \cite{onDL} shows that the E2E paradigm experiences performance degradation, even when only 10\% of unseen data is omitted from the training set. Considering the aforementioned challenges, there is a research gap in the E2E learning-based data transmission with large block length over an unknown channel model, while concurrently addressing the problem of local optimization arising from the non-differentiable channel function. Hence, it demands an innovative E2E communication system capable of adaptive training on various channel models without prior knowledge of explicit channel effects. Additionally, the proposed scheme should demonstrate the ability to transmit long sequences of bit-stream without compromising performance.

Given the above, this article introduces a novel deep reinforcement learning (DRL)-based E2E communication system empowered by the deep deterministic policy gradient (DDPG) algorithm, where both the transmitter and the receiver can be jointly trained over an unknown channel. This approach leverages the loss information obtained from the receiver as the reward signal and directly applies it to optimize the transmitter which acts as the agent in the DDPG algorithm. Additionally, a CNN-based architecture is developed to accommodate input and output messages with large block lengths. The proposed scheme exhibits notable improvement in BLER performance compared to existing solutions under complex wireless communication scenarios, particularly when operating with large block lengths (e.g., 256 information bits). Our major contributions are summarized as follows.

\begin{itemize}
    \item We develop a DDPG-based E2E communication system, which can jointly optimize the transmitter and the receiver without prior knowledge of channel models. In particular, the proposed DDPG employs actor and critic networks to encode information at the transmitter before transmitting it over an unknown channel by leveraging the training loss feedback from the receiver. As the loss feedback implicitly captures the training process of the receiver and the channel conditions, the proposed algorithm can help to jointly train the transmitter and the receiver, thus significantly improving the communication performance of the whole system.
    
    \item To address the issue of dimensionality, CNN is employed to extend the block length of the system from 8 bits to 256 bits. This approach allows the transmitter to process high dimensional input data, and the receiver then learns to recover the encoded signals through our proposed CNN.
    
    \item Through extensive simulations, we demonstrate that the proposed DDPG-based E2E communication system can achieve better BLER performance and convergence rate compared to state-of-the-art solutions over complex wireless communication channels, e.g., Rician, Rayleigh, and 3GPP channels.
\end{itemize}

The rest of the paper is organized as follows. The related works will be reviewed in Section~\ref{sec:Related_Works}. In Section~\ref{sec:Problem_Formulation}, the modeling of E2E communication systems is presented. In Section~\ref{sec:DDPG_E2E}, the proposed DDPG-based E2E system is explained in detail. Then, the training process of the E2E system is presented in Section~\ref{sec:Training_E2E}. The simulation results are presented in Section~\ref{sec:evaluation}. Finally, Section~\ref{sec:conclusion} concludes our paper.

\underline{\textit{Notation:}} The notations used in this paper are listed in the following. Lowercase letters $\bm{m}$ and $\hat{\bm{m}}$ denote the original message and the estimated message. $\bm{x}$ denotes the encoded signal, and $\bm{y}$ denotes the received signal. ${\mathbb{C}}^{n}$ denotes the complex values of $n$ dimension. $\bm{s}_{t}$, $\bm{a}_{t}$ and $r_{t}$ denote the state, action and reward at time step $t$, respectively. $\mu$ denotes the policy function mapping state $\bm{s}$ to action $\bm{a}$. $Q(\bm{s},\bm{a})$ denotes the Q-value which measures the expected cumulative reward that an agent can obtain by taking a particular action $\bm{a}$ in a particular state $\bm{s}$ following a policy function $\mu$.

\section{Related Works}
\label{sec:Related_Works}

\subsection{Deep Learning in End-to-End Communication Systems}
In recent years, different DL techniques have been applied to E2E communication systems to solve the unknown channel problem. The study is pioneered in~\cite{overAir}, a two-step training scheme using transfer learning is developed, where the E2E communication system is trained on an assumed stochastic channel model that is close to the actual channel effects at the first stage. Subsequently, the receiver is then finetuned based on the difference between the assumed stochastic channel model and the practical channel model at the second stage. This proposed scheme showcases the possibility of conducting over-the-air data transmission entirely carried out by NNs. However, the transmitter cannot be optimized during the finetuning of the receiver, which hinders the overall learning of the E2E communication system. There are some existing works studying to improve the two-step training approach, which can be classified into two main categories, i.e., channel imitation approaches and receiver-aided approaches.

\subsection{Channel Imitation Schemes for End-to-End Communication Systems}

In order to learn the channel effects in practical channel models and address the challenges of local optimization caused by the BP algorithm during the auto-encoder training~\cite{BP}, the channel imitation scheme using generative adversarial network (GAN) is widely studied~\cite{goodfellow2020generative}. In~\cite{GANsmall}, a novel conditional GAN-based channel agnostic E2E communication system is developed to adaptively learn the channel effects of different types of channel models. More specifically, the generator is trained to approximate the conditional data distribution of the practical channel effects, and the discriminator is trained to distinguish the real signals from the generated fake signals. By conditioning the generator and the discriminator on the channel, the system can learn to adapt to a variety of channel conditions, even if the channel characteristics are unknown beforehand. This channel agnostic E2E system empowers the generator to serve as a bridge connecting the transmitter and the receiver. This enables the BP of error gradients from the receiver to the transmitter, ultimately facilitating the overall learning of the transceiver. This conditional GAN-based channel imitation scheme is further enhanced in~\cite{GAN}, where the system can not only adapt to various channel models by learning their characteristics but can also transmit long sequences of information bits. In particular, this proposed system has a significantly extended block length, from 8 information bits to 128 information bits, without much degradation in BLER performance. However, the GAN-based schemes face challenges that they struggle to capture complex channel distributions and require extensive training data, leading to poor generalization in real-world channel conditions.

A residual-aided GAN (RA-GAN) based E2E communication system is proposed to solve the above challenges by applying residual neural network (ResNet)~\cite{resnet} to the generator to enable better generation performance for the received signals and adding $\bm{l}_2$ regularizer to punish large weights in E2E system~\cite{ra-gan}. However, this scheme experiments only on a block length of 4 bits, which falls considerably short of the practical scenarios for data transmission. Despite the effectiveness of the aforementioned channel imitation schemes in addressing the local optimization issue associated with the BP algorithm, GAN-based solutions still show limited performance with large block lengths. This limitation is particularly evident in more complex wireless channels, such as the Rayleigh fading channel. As such, further improvements are required to address these challenges and enhance the system's performance.

\subsection{Receiver-Aided Schemes for End-to-End Communication Systems}

Recently, DRL has demonstrated its great potential in wireless communications and networking \cite{hoang2023deep}, particularly in the field of E2E communication systems design. Apart from the channel imitation solutions, RL techniques have been emerging as a promising tool in receiver-aided approaches, enabling data transmission without the need for explicit channel modelling or CSI~\cite{raj2018backpropagating}. In particular, the simultaneous perturbation stochastic approximation technique is used in~\cite{without} to train the transmitter, where the transmitter is optimized by utilizing the loss value obtained from the receiver. In~\cite{model_free}, a combination of RL and supervised learning techniques is employed to optimize both the transmitter and the receiver. In~\cite{noisy_feedback}, the authors extend these works by considering a noisy feedback loop. Particularly, the proposed system with a noisy feedback link demonstrates identical performance with the system with a perfect feedback link. In summary, the aforementioned receiver-aided solutions circumvent the unknown channel issue and local optimization problem arising from BP by using the loss value obtained from the receiver. However, the biggest issue of the receiver-aided solutions is that the variance of loss value from the receiver will scale with the increasing number of channel uses and the increasing block length of the input message, which leads to significant performance degradation and slow convergence. In addition, the training of communication systems over complex wireless communication channels faces challenges due to the relatively long coherence time where only small portions of CSI are learned, which leads to poor sample efficiency and also slow convergence. Given the above disadvantages of the channel imitation approaches and receiver-aided approaches, there is currently no effective solution that can transmit data with large block lengths (e.g., 256 information bits) over an unknown channel without compromising the BLER performance.

\subsection{CNN-based Architecture for Binary Encoding}

CNNs are better suited than Multi-layer perceptrons (MLPs) for handling large block lengths with binary encoding because they can effectively capture local patterns through convolutional layers, which enhances their ability to recognize bit sequences. Additionally, CNNs benefit from parameter efficiency due to weight sharing, avoiding the parameter explosion issue that can occur with MLPs when dealing with large block sizes \cite{garcia2018survey}. In order to tackle the encoding of source information with large block length, the binary encoding is adopted rather than the one-hot encoding in many existing works including \cite{ye2019circular}, \cite{GAN} and \cite{dorner2022learning}. These works introduce novel CNN architectures to mitigate the dimensionality explosion associated with larger block sizes. Specifically, the first two works, \cite{ye2019circular} and \cite{GAN}, focus on using CNNs for binary encoding with large block lengths, which requires explicit signal processing including channel estimation and channel equalization. In addition, the authors in \cite{jiang2019turbo} introduce an innovative Turbo Autoencoder (TAE) architecture that combines the strength of Turbo coding and CNNs to effectively model the encoder and decoder processes in communication systems. In particular, the proposed TAE exhibits superior bit error rate performance under low SNR conditions when compared to the conventional coding schemes, including LDPC and polar codes. Moreover, it demonstrates significant performance improvement for large block sizes over the classical Convolutional-Autoencoder model presented in \cite{jointtrans}. In contrast, \cite{dorner2022learning} extends the use of CNNs to a fully differentiable end-to-end system that includes detection, synchronization, equalization, and decoding. In our work, we develop a CNN architecture inspired by \cite{ye2019circular} to tackle the challenges of large block sizes. Therefore, the synchronization and channel estimation are not considered in the work. With the proposed CNN, we effectively train an E2E communication system for longer block sizes without explicit information about channel distributions in advance.


\section{System Model}
\label{sec:Problem_Formulation}

In this section, we will introduce the basic concepts of E2E communication systems. As shown in Fig.~\ref{fig:architecture}(b), an auto-encoder-based E2E communication system consists of a transmitter, a receiver, and a mathematically formulated channel, where the transmitter and the receiver are represented as two independent NNs. The transmitter NN aims to encode the original binary message $\bm{m}$ to the encoded signal $\bm{x} \in {\mathbb{C}}^{n}$, where ${\mathbb{C}}^{n}$ denotes that the encoded signal $\bm{x}$ is presented as complex numbers, making $n$ discrete uses of channel. The encoded signal is then sent via the channel to the receiver. The received signal after channel distortion is denoted by $\bm{y} \in {\mathbb{C}}^{n}$. The receiver aims to decode the received signals and recover the original binary message $\bm{m}$. The implementations of the E2E communication system can be illustrated as the sequence of three individual functions, which can be expressed as:
\begin{equation}
\hat{\bm{m}}=f_\mathrm{D}(f_\mathrm{h}(f_\mathrm{E}(\bm{m}; \bm{\theta}_\mathrm{E})); \bm{\theta}_\mathrm{D}),
\label{eq:1}
\end{equation}
where $f_\mathrm{E}$ represents the encoder function mapping the original binary message to the encoded signal, i.e., $\bm{x}=f_\mathrm{E}(\bm{m}; \bm{\theta}_\mathrm{E})$, and $\bm{\theta}_\mathrm{E}$ denotes the trainable parameters of the transmitter NN. $f_\mathrm{h}$ represents the channel impairments with channel realization $h$, i.e., $\bm{y}=f_\mathrm{h}(\bm{x})$. $f_\mathrm{D}$ represents the decoder function that aims to recover the received signals to the estimated message, i.e., $\hat{\bm{m}}=f_\mathrm{D}(\bm{y}; \bm{\theta}_\mathrm{D})$, and $\bm{\theta}_\mathrm{D}$ denotes the trainable weights of the receiver NN. The goal of the E2E communication system is to jointly optimize both the transmitter NN and the receiver NN by minimizing the E2E loss function, denoted by $L=\mathcal{L}(\bm{m}, \hat{\bm{m}})$. The loss function $\mathcal{L}(\bm{m},\hat{\bm{m}})$ is regarded as the objective function which measures the distance between the original message $\bm{m}$ and the estimated message $\hat{\bm{m}}$ for such E2E framework.

The E2E learning approach considers the entire system as an auto-encoder and trains the NNs of the transceivers in a supervised manner. The training process involves calculating the error gradients of each layer by utilizing the BP algorithm. To enable the BP of the error gradients, it is necessary to ensure that the loss function with respect to the trainable weights at each layer of the NN is differentiable. The computed error gradients subsequently update the weights and bias at each layer, with the goal of minimizing the overall E2E loss. Therefore, the E2E learning approach entails a differentiable channel function beforehand to facilitate the BP of error gradients from the receiver to the transmitter. This enables the joint optimization of the entire E2E communication system. However, while the E2E learning paradigm demonstrates efficiency in simple channel models, its performance becomes notably constrained when confronted by more complex communication scenarios. The current limitations of the E2E learning paradigm can be summarized as follows. First, the primary limitation of the E2E learning method is its reliance on an explicit differentiable channel model, which is often impractical due to real-world communication challenges. Second, the non-differentiable channel model potentially hinders the gradient-based optimization, preventing the overall learning of the system. Third, obtaining real-time CSI is difficult in practice, undermining the validity of assumed channel functions. Furthermore, the existing literature primarily focuses on research with short block lengths, and there is a lack of evidence regarding the performance of solutions for block lengths exceeding 100 bits.

This paper introduces a DDPG-based E2E communication system aimed at addressing the challenges posed by non-differentiable channel models and imperfect CSI while improving the performance in terms of data transmission with long block lengths. By leveraging the DDPG algorithm, the loss value of the receiver NN serves as feedback information for updating the transmitter NN. Hence, the transmitter NN and the receiver NN can be implicitly and jointly optimized without requiring prior knowledge of channel models. The following sections will present a comprehensive analysis of our proposed DDPG-based E2E solution.

\section{Deep Deterministic Policy Gradient For End-to-End Communication Systems}
\label{sec:DDPG_E2E}

In this section, we first introduce the basic concepts of DRL. Then, the overview of the proposed DDPG-based E2E communication system and the detailed network architectures are presented.

\subsection{Deep Reinforcement Learning Basics}

\begin{figure*}[!]
	\centering
        \includegraphics[scale=0.45]{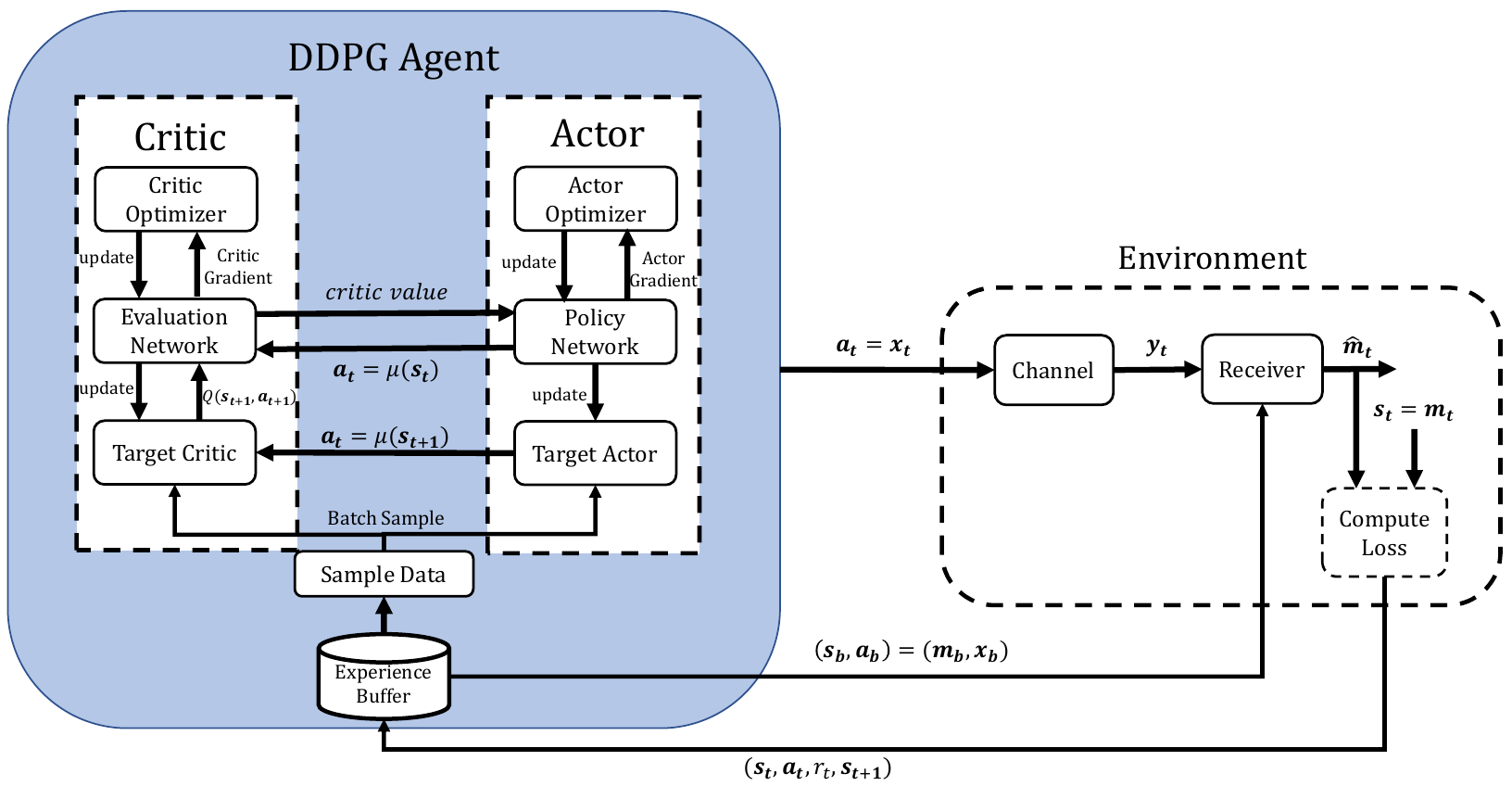}
	\caption{Overview of DDPG-based E2E communication system.}
	\label{fig:DDPG}
\end{figure*}

In the field of reinforcement learning, a typical system comprises an agent and an environment. The primary objective of an RL system is to train the agent in a manner that allows it to interact with an unknown environment, gather valuable experiences, and maximize cumulative rewards through continuous exploration. In particular, at time step $t$, the agent observes the environment, i.e., state $\bm{s}_{t}$, and makes action $\bm{a}_{t}$ based on its current policy $\pi$. After performing action $\bm{a}_{t}$, the agent observes immediate reward $r_{t}$ and next state $\bm{s}_{t+1}$. For a given policy $\pi$, the action-value function, i.e., Q-function, which measures the expected return for a given state-action pair, can be mathematically expressed as:
\begin{equation}
Q_{\pi}(\bm{s},\bm{a})=\mathbb{E}_{\pi}[G_{t}|\bm{s}_{t}=\bm{s},\bm{a}_{t}=\bm{a}],
\label{eq:Q}
\end{equation}
where $G_{t} = \sum^{\infty}_{t=1}\gamma^{t}r_{t}$ denotes the cumulative discounted reward for one trajectory, and $\gamma \in (0, 1]$ represents the discount factor which determines the importance of future rewards relative to immediate rewards. In Q-learning, the Q-value update rule for learning the optimal policy is given by:
\begin{equation}
    \begin{aligned}
            &Q(\bm{s}_{t},\bm{a}_{t}) {\leftarrow} Q(\bm{s}_{t},\bm{a}_{t})\\
        &+{\alpha}[R(\bm{s}_{t},\bm{a}_{t})+{\gamma}(\underset{a_{t+1}}{\max}Q(\bm{s}_{t+1},\bm{a}_{t+1}))-Q(\bm{s}_{t},\bm{a}_{t})],
    \end{aligned}
\end{equation}  
where $R(\bm{s}_{t},\bm{a}_{t})$ denotes the immediate reward given state $\bm{s}_t$ and action $\bm{a}_t$. $\bm{s}_{t+1}$ denotes the next state that the agent transits to after taking action $\bm{a}_t$ in state $\bm{s}_t$, $\bm{a}_{t+1}$ is the next action that the agent takes at state $\bm{s}_{t+1}$, and $\alpha$ is the learning rate which determines how much weights is given to the new information. This algorithm iteratively improves the policy by updating the Q-value and choosing the action with the highest Q-value at each state, which aggressively targets the local optimum for every decision during exploration. Hence, the $\epsilon-greedy$ mechanism \cite{greedy}, an exploration strategy, is usually used to balance the exploration and exploitation during the learning process. It uses small probability $\epsilon$ to select a random action to encourage exploration and selects the action with the highest estimated value to exploit the current knowledge with a probability of $1-\epsilon$. The update of the Q-function using $\epsilon-greedy$ algorithm \cite{greedy} can be expressed as:

\begin{align}
 \begin{split}
Q_{\pi}(\bm{s}, \pi'(\bm{s})) = \frac{\epsilon}{|A|}\sum_{\bm{a}\in\mathbb{A}} Q_{\pi}(\bm{s}, \bm{a}) + (1 - \epsilon) \max_{\bm{a} \in \mathbb{A}} Q_{\pi}(\bm{s}, \bm{a}),
\label{eq:Value}
 \end{split}
\end{align}

\noindent where $\epsilon$ is the probability of making an exploratory action at every time step. In practice, the exploration probability $\epsilon$ is typically decreased gradually as the algorithm converges, allowing the policy to exploit the learned information more often. $\pi'$ refers to the target policy, which selects actions based on the balance between exploration and exploitation, while $\pi$ refers to the current policy being evaluated.

To further enhance the convergence of Q-learning, the concept of deep Q-learning has been introduced, replacing the traditional Q-table with a NN, known as deep Q-network (DQN)~\cite{DQN}. DQN harnesses the computational power of NN to address regression problems by taking the state-action pair as input and approximating the corresponding Q-value through predictive modelling. The primary advantage of deep Q-learning over conventional Q-learning lies in its efficiency, particularly when dealing with complex tasks in which Q-learning requires a very long time to learn or even cannot obtain the optimal solution. Furthermore, deep Q-learning introduces an innovative technique known as experience replay \cite{ExperienceReplay}. This method involves storing accumulated information in a buffer, enabling the agent to sample from past experiences to optimize the policy network. The experience replay effectively utilizes the past cumulative experience and mitigates the issues of data correlation in the observation sequences by random sampling. However, deep Q-learning encounters several limitations. The most critical issue of deep Q-learning is that it cannot directly operate in continuous actions space, whereas many real-world applications precisely require continuous actions space. For instance, practical communication systems are often involved with continuous, time-varying encoded signals and can be difficult to measure by discrete states. Another challenge is the frequent update of the target network, causing the behaviour network continuously ''chasing'' a moving target \cite{inbook}. To address this, we propose a DDPG-based approach for E2E communication systems. This method effectively manages continuous action spaces and allows joint optimization for the transceivers without requiring prior channel information.

\subsection{End-to-End Communication Systems using DDPG}

DDPG, or Deep Deterministic Policy Gradient, represents an off-policy actor-critic algorithm \cite{actorcritic} that effectively combines the advantages of deep Q-learning and policy gradient \cite{ddpg}. The proposed DDPG-based system encompasses several key features: (i) it employs a deterministic policy for network optimizations, allowing the agent to directly operate on the continuous action space, (ii) it follows an actor-critic architecture, where the actor network generates the deterministic action based on the current state, and the critic network evaluates the performance of the current action, and (iii) to ensure stable training, the DDPG approach uses two target networks - one for the actor and another for the critic. These target networks serve as time-delayed copies of their original networks that slowly track the behaviour of the behaviour networks. This off-policy design significantly contributes to the overall stability of the training process. 

The schematic overview of the proposed DDPG-based E2E communication system is visualized in Fig.~\ref{fig:DDPG}. The observation state corresponds to the input message at time step \( t \), i.e., \( \bm{s}_{t} = \bm{m}_{t} \). The action represents the encoded signal, i.e., \( \bm{a}_{t} = \bm{x}_{t} \). The received signal is denoted by \( \bm{y}_{t} \), while \( \hat{\bm{m}} \) signifies the estimated output message. Additionally, \( \bm{s}_\mathrm{b} \) and \( \bm{a}_\mathrm{b} \) represent the randomly sampled state and action batches from the experience buffer used for training the transmitter and receiver. At the core of the system, the environment comprises two key elements, including the channel model and the receiver model. The DDPG agent encompasses the actor and the critic networks. The actor network, serving as the transmitter model, maps the input message $\bm{m}$ to the encoded signal $\bm{x}$ and transmits it through the unknown channel. The channel model introduces distortions to the encoded signal $\bm{x}$ before sending it to the receiver model. The receiver model then decodes and recovers the distorted signal to produce the estimated message $\hat{\bm{m}}$. The binary cross-entropy loss \cite{bce} is employed to measure the distance between the original information $\bm{m}$ and the estimated message $\hat{\bm{m}}$. The immediate reward after performing an action is the opposite of the computed loss since the loss negatively reflects the model performance, which can be expressed as:
\begin{equation}
\label{eq:rewardfunction}
    \begin{aligned}
        r_{t}=&-\sum\limits_{k=1}^{K} \left[ (\bm{s}_{t})_{k} \log \left(f_{\bm{\theta}_\mathrm{R}} \left(\mu(\bm{s}_{t} \mid \bm{\theta}_\mathrm{T}) \right) \right)_{k} \right. \\
        &\left. + \left(1 - (\bm{s}_{t})_{k} \right) \log \left(1 - f_{\bm{\theta}_\mathrm{R}} \left(\mu(\bm{s}_{t} \mid \bm{\theta}_\mathrm{T}) \right) \right)_{k} \right],
    \end{aligned}
\end{equation}
\noindent where $\bm{\theta}_\mathrm{R}$ denotes the weights of the receiver network, $\bm{\theta}_\mathrm{T}$ denotes the weights of the transmitter network, and $\mu(\bm{s}_{t}|\bm{\theta}_\mathrm{T})$ denotes the action taken by policy $\mu$ at state $\bm{s}_{t}$ based on the transmitter weights $\bm{\theta}_\mathrm{T}$. $f_{\bm{\theta}_\mathrm{R}}$ represents the process at the receiver model mapping the distorted signal to the estimated message, and $K$ denotes the length of the input message. The reward $r_{t}$ is the feedback signal from the receiver for updating both the actor and the critic networks at the agent. $k$ denotes the number of indices in the binary message. Due to the dynamic nature of wireless communications, the transmitted bits are typically independent and uncorrelated, resembling a multi-armed bandit problem. To align with the randomness between the signals, we assume that the adjacent states are uncorrelated and independent of each other. A similar approach has been implemented in \cite{model_free}, where the author also assumes the uncorrelated input batches. DRL enables the system to optimize the transceivers by leveraging an exploration mechanism, thereby facilitating adaptation from known to unknown channel conditions. Since the next state $\bm{s}_{t+1}$ is randomly generated and is independent of the current state $\bm{s}_{t}$, the return of the Bellman function should focus more on the immediate reward rather than the future value. For that, the discount factor $\gamma$ is set at 0.01 to eliminate the correlated effect between the adjacent states \cite{duckworth2023reinforcement}.

\subsection{Network Architectures of Transmitter, Receiver and Critic}

\begin{figure*}[!]
	\centering
	\begin{subfigure}[b]{0.33\textwidth}
		\centering
		\includegraphics[width=\textwidth]{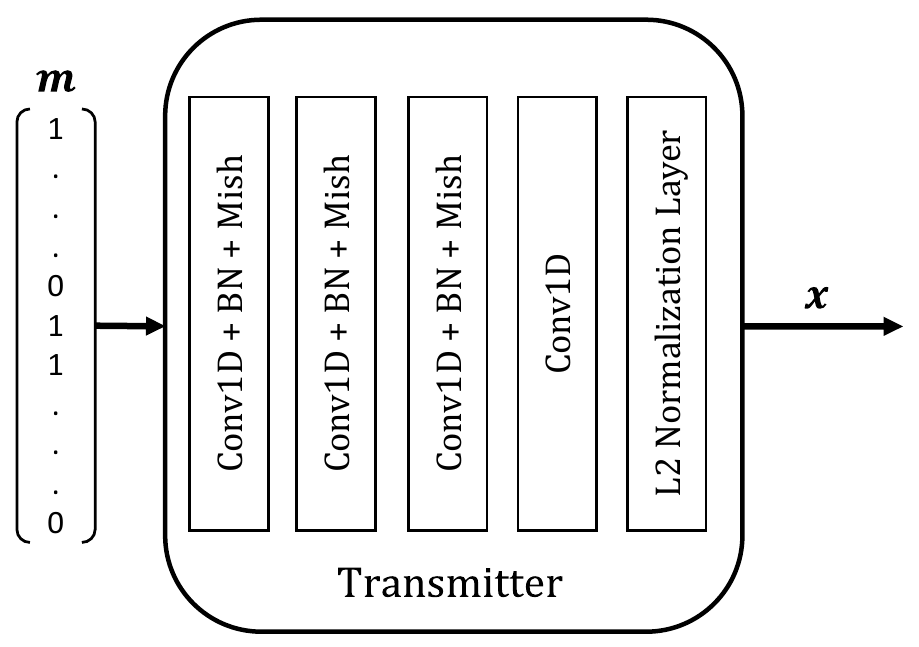}
		\caption{}
	\end{subfigure}
	\hspace{1cm}
	\begin{subfigure}[b]{0.4\textwidth}
		\centering
		\includegraphics[width=\textwidth]{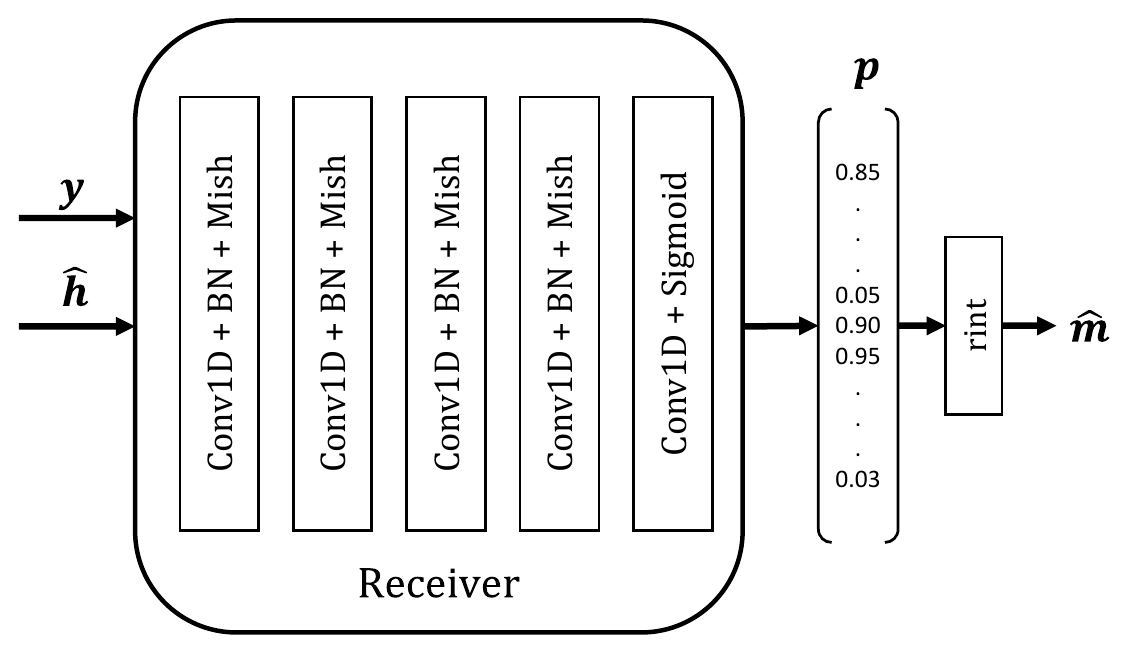}
		\caption{}
	\end{subfigure}
	\hfill
	\begin{subfigure}[b]{0.4\textwidth}
		\centering
		\includegraphics[width=\textwidth]{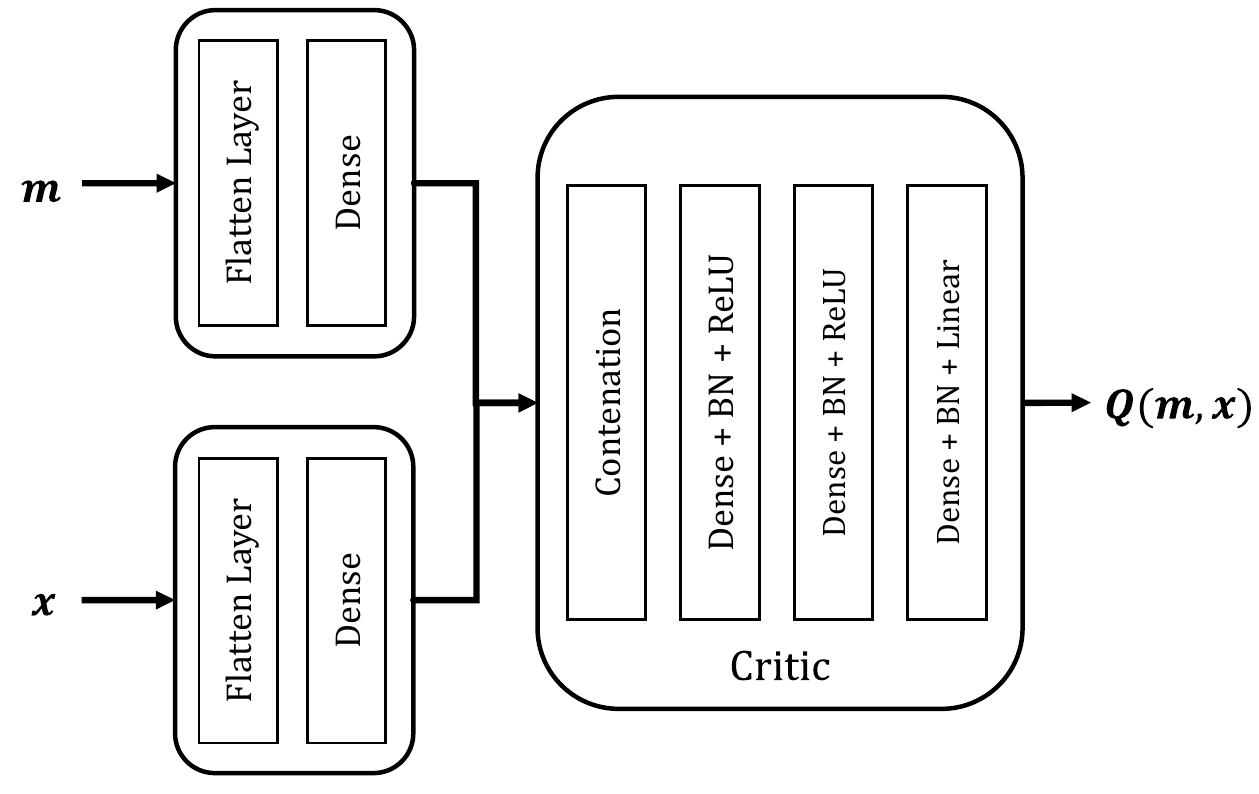}
		\caption{}
	\end{subfigure}
	\caption{Architectures of (a) transmitter, (b) receiver, and (c) critic networks.}
	\label{fig:network_arch}
\end{figure*}

Fig.~\ref{fig:network_arch} provides an illustration of the network architectures of the transmitter, the receiver, and the critic models. In our proposed solution, CNNs are employed to facilitate the curse of dimensionality. The detailed parameter configurations for the network architectures of the transmitter and the receiver are presented in Table \ref{Tab:ArchCNN}, with $K$ denoting the block length. At the transmitter, the original message $\bm{m}$ is a binary number vector of size $K$, i.e., $\bm{m}\in\{0, 1\}^{K}$. This message $\bm{m}$ serves as the input to the transmitter. More specifically, the transmitter architecture depicted in {Fig.~\ref{fig:network_arch}(a) comprises four one-Dimentional convolutional (Conv1D) layers using circular padding techniques \cite{ye2019circular}. Subsequently, an L2 normalization layer \cite{l2} is applied to normalize the total power of the encoded signal, ensuring $||\bm{x}||^{2} \leqslant n$. The initial convolutional layer employs 256 filters, followed by a batch normalization (BN) layer~\cite{BatchNormalization} and the Mish activation function \cite{misra2019mish}.} The number of filters progressively decreases to 2 for reshaping the dimension of the encoded signal. The resulting encoded signal would be a two-dimensional vector with a length of $K$, representing the complex-valued symbols.


At the receiver, the estimated channel response $\hat{\bm{h}}$ and the received signal $\bm{y}$ are combined and served as the inputs of the receiver network. The receiver network is composed of three Conv1D layers, followed by an additional Conv1D layer with the Sigmoid activation function \cite{sigmoid}. The latter layer is responsible for generating the estimates of the probabilities associated with all possible output messages $\bm{p}$. This output message vector $\bm{p}$ will be rounded to a binary number vector, i.e., the estimated message $\hat{\bm{m}}$. The block \emph{rint} shown in Fig.~\ref{fig:network_arch}(b) denotes that probabilities of the output vector are rounded to integers. Fig.~\ref{fig:network_arch}(c) shows the network architecture of the critic model. The critic network takes both the message $\bm{m}$ and the encoded signal $\bm{x}$ as the input and generates $Q(\bm{m},\bm{x})$ as the critic value, where each input follows with a Flatten layer and a dense layer. The aligned inputs are then concatenated to a one-dimension vector with a length of 256. The concatenated input is then passed through three sequential layers, each consisting of a fully connected (Dense) layer followed by BN and Rectified Linear Unit (ReLU) activation function~\cite{ReLU}. The first two Dense layers consist of 256 neurons each, while the final Dense layer has a single neuron output for generating the critic value. The final Dense layer is followed by BN and a linear activation function.

\begin{table}[h]
\centering
\begin{tabular}{|c|c|c|}
 \hline
 Type of layer & Kernel size & Output size \\ 
 \hline
 \hline
 \multicolumn{3}{|c|}{\textbf{Transmitter}} \\
 \hline
 Input & Input layer & $K \times 1$\\
 \hline
 Conv1D+BN+Mish & 5 & $K \times 256$\\
 \hline
 Conv1D+BN+Mish & 3 & $K \times 128$\\
 \hline
 Conv1D+BN+Mish & 3 & $K \times 64$\\
 \hline
 Conv1D & 3 & $K \times 2$\\
 \hline
 L2 Normalization & Power normalization & $K \times 2$\\
 \hline
 \hline
 \multicolumn{3}{|c|}{\textbf{Receiver}}\\
 \hline
 Input & Input layer & $K \times 4$\\
 \hline
 Conv1D+BN+Mish & 5 & $K \times 256$\\
 \hline
 Conv1D+BN+Mish & 5 & $K \times 128$\\
 \hline
 Conv1D+BN+Mish & 5 & $K \times 64$\\
 \hline
  Conv1D+BN+Mish & 5 & $K \times 32$\\
 \hline
 Conv1D+Sigmoid & 3 & $K \times 1$\\
 \hline
 \hline
 \multicolumn{3}{|c|}{\textbf{Critic}}\\
 \hline
 Input1 & State Input layer & $K$\\
 \hline
 Input2 & Action Input layer & $K \times 2$\\
 \hline
 Flatten1 & State Flatten layer & $K$\\
 \hline
 Flatten2 & Action Flatten layer & $2K$\\
 \hline
 Dense1+ReLU & N/A & 128\\
 \hline 
 Dense2+ReLU & N/A & 128\\
 \hline
 Concatenate & N/A & 256\\
 \hline
 Dense+BN+ReLU & N/A & 256\\
 \hline
 Dense+BN+ReLU & N/A & 256\\
 \hline
 Dense+BN+Linear & N/A & 1\\
 \hline
\end{tabular}
\caption{Architectures of transmitter, receiver, and critic networks.}
\label{Tab:ArchCNN}
\end{table}

\section{Training of DDPG-based End-to-end Communication Systems}
\label{sec:Training_E2E}
In this section, we will first introduce the overall training process. Then, the detailed training implementations of the transmitter and the receiver will be presented.

\subsection{Training Process Overview}

\begin{algorithm}[!]
\caption{DDPG for E2E Communication Systems}
\label{alg:algo1}
\begin{algorithmic}[1]
    \State \multiline{Randomly initialize actor network ${\bf \mu}(\bm{s} \mid \bm{\theta}^\mu)$ and critic network ${Q}(\bm{s},\bm{a} \mid \bm{\theta}^Q)$}
    \State{Initialize target network ${Q}'$ and ${\mu}'$ with the same weights}
    \State{Initialize Buffer with capacity of $C$ and batch size $B$}
    \For{\textit{episode = 1 to E}}
    \State{Initialize a random state for the input message  $\bm{s}_t$=$\bm{m}$}
    \State{Initialize the episodic reward as zero}
    \For{\textit{step = 1 to T}}
    \State{\multiline{Assign next state from last iteration to current state: $\bm{s}_{t}=\bm{s}_{t+1}$}}
    \State \multiline{Select the action (encoded signal) from actor network: $\bm{x} = \bm{a}_t = \mu(\bm{s} \mid \bm{\theta}^\mu)$}
    \State {Feed the state $\bm{s}_t$ and action $\bm{a}_t$ to the environment}
    \State \multiline{Observe the new state $\bm{s}_{t+1}$, reward $r_t$}
    \State \multiline{Store the transition $(\bm{s}_{t},\bm{a}_{t},r_t,\bm{s}_{t+1})$ to Buffer}
    \State \textsc{TrainAgent}
    \State \textsc{TrainRx}
    \EndFor
    \EndFor
\end{algorithmic}
\end{algorithm}

\begin{algorithm}[!]
\caption{Training of the Agent (Actor, Critic, Target Actor and Target Critic)}
\label{alg:algo2}
\begin{algorithmic}[1]
    \Function{TrainAgent()}{}
        \State \textbf{Input:} \multiline{Sample a batch of $B$ transitions $(\bm{s}_{i},\bm{a}_{i},r_i,\bm{s}_{i+1})$} 
        \State \multiline{Set $\bm{y}_{i}^\text{target}=r_{i}+\gamma{Q(\bm{s}_{i+1},\bm{a}_{i+1})}$}
        \State \multiline{Update critic Network by minimizing the loss:}
    \begin{equation}
        \mathcal{L}_{\bm{\theta}^\mathrm{\mu}}=\frac{1}{B}\sum_{i}(\bm{y}_{i}^\text{target}-Q(\bm{s}_{i},\bm{a}_{i}|\bm{\theta}^{Q}))^{2} \nonumber
    \end{equation}
    \State \multiline{Update actor Network using sampled policy gradient:}
    \begin{equation}
        \begin{aligned}
        {{\nabla }_{{{\bm{\theta} }^{\mu }}}}J & \approx \frac{1}{B}\sum\limits_{i}{{{\nabla }_{\bm{a}}}Q(\bm{s},\bm{a}|{{\bm{\theta} }^{Q}}){{|}_{\bm{s}={\bm{s}_{i}},\bm{a}=\mu ({\bm{s}_{i}})}}{{\nabla }_{{{\bm{\theta} }^{\mu }}}}\mu (\bm{s}|{{\bm{\theta} }^{\mu }}){{|}_{{\bm{s}=\bm{s}_{i}}}}} \nonumber
    \end{aligned}
    \end{equation}
    \State{Update Target networks for both actor and critic:}
    \begin{equation}
        \begin{aligned}
            \bm{\theta}^{Q'}&=\tau\bm{\theta}^{Q}+(1-\tau)\bm{\theta}^{Q'} \\
            \bm{\theta}^{\mu'}&=\tau\bm{\theta}^{\mu}+(1-\tau)\bm{\theta}^{\mu'} \nonumber
        \end{aligned}
    \end{equation}
    \EndFunction
\end{algorithmic}
\end{algorithm}

\begin{algorithm}[!]
\caption{Training of the Receiver}
\label{alg:algo3}
\begin{algorithmic}[1]
    \Function{TrainRx}{}    
        \State \textbf{Input:} Sample state-action batches $(\bm{s}_{b},\bm{a}_{b})$ with a batch size of $B$
        \State Perform supervised learning with the sampled state batch as the input data, and the sampled action batch as the label
        \State 
        \begin{equation}
            f_{\bm{\theta}_\mathrm{R}}({\bm{a}_b}) = {\hat{\bm{s}}_b} \nonumber
        \end{equation}
        \State Calculate the binary cross-entropy loss for the receiver
        \begin{equation*}
            \begin{aligned}
                \mathcal{L}_{\bm{\theta}_\mathrm{R}} = - \frac{1}{K} \sum_{k=1}^{K} {(\bm{s}_b)}_k \log({(\hat{\bm{s}}_b)}_k) 
                + (1 - {(\bm{s}_b)}_k) \log(1 - {(\hat{\bm{s}}_b)}_k)
            \end{aligned}
        \end{equation*}
        \State Update the model parameters using the computed loss
    \EndFunction
\end{algorithmic}
\end{algorithm}

The diagram overview of the training process is depicted in Fig. \ref{fig:DDPG} and the pseudocode of the whole training process is presented in Algorithm \ref{alg:algo1}. At the initial phase, both the actor and the critic networks are initialized with random parameters $\bm{\theta}^{\mu}$ and $\bm{\theta}^{Q}$, respectively. Two target networks, denoted as $\mu'$ and $Q'$ are initialized with identical parameters for the initial equilibrium. A replay buffer with a capacity of $C$ is initialized to store the cumulative information obtained from the environment. During the training phase, a random observation state is generated at the beginning of each episode, which can be represented as $\bm{s}_t=\bm{m}$. At every time step within each episode, the actor network maps the current observation state $\bm{s}_{t}$ to the action, i.e., encoded signal $\bm{a}_t=\bm{x}$. The chosen action is then conveyed to the environment, where the observation state is considered as the label of the estimated output for calculating the loss value. Subsequently, the environment will then return the next observation state $\bm{s}_{t+1}$, the calculated reward $r_{t}$ and the $done$ information as feedback to the agent, where the $done$ information is a Boolean variable that determines whether to terminate the training of current episode. A collection of state, action, reward, and next state will be stored in the experience buffer of the agent.Once the data is collected, it starts to train the agent as depicted in Algorithm \ref{alg:algo2}. A mini-batch of experiences is randomly sampled from the replay buffer. For each state-action pair in the mini-batch, the target Q-value $\bm{y}_{i}^\text{target}$ is computed using the following equation:
\begin{equation}
\begin{aligned}
\bm{y}_{i}^\text{target} &= r_{i} + \gamma Q(\bm{s}_{i+1}, \bm{a}_{i+1}) \\
&= r_{i} + \gamma Q(\bm{s}_{i+1}, \mu(\bm{s}_{i+1} \mid \bm{\theta}^{\mu'}) \mid \bm{\theta}^{Q'}).
\end{aligned}
\end{equation}

The critic network is then trained to minimize the mean square error between the predicted and target Q-values using the mini-batch of experiences. The loss function between the estimated and target Q-values can be expressed as:
\begin{equation}
\mathcal{L}_{\bm{\theta}^\mathrm{\mu}}=\frac{1}{B}\sum_{i}(\bm{y}_{i}^\text{target}-Q(\bm{s}_{i},\bm{a}_{i}|\bm{\theta}^{Q}))^{2}, 
\end{equation}
where $Q(\bm{s}_{i},\bm{a}_{i}|\bm{\theta}^{Q})$ denotes the predicted Q-value generated by the behaviour networks. In the process of optimizing the actor and the critic networks, the objective is to iteratively update these networks to maximize the expected return, i.e., the expected value of the Q-function. The expected value of the Q-function can be mathematically represented as:
\begin{equation}
    J(\bm{\theta}) = \mathbb{E}[Q(\bm{s},\bm{a})|_{\bm{s}=\bm{s}_{i},\bm{a}_{i}=\mu(\bm{s}_{i})}],
\end{equation}
where $\bm{s}_{i}$ denotes the current state of the environment, and $\bm{a}_{i}$ denotes the current action selected by the policy function, i.e., $\bm{a}_{i}=\mu(\bm{s}_{i})$. In the optimization process of the actor network, a crucial step is to calculate the gradient of the expected Q-value concerning the policy parameters $\bm{\theta}_\mathrm{\mu}$. This enables the actor network to learn more favorable actions in different states. The gradient of the expected Q-value with respect to the policy parameter is mathematically formulated as: 
\begin{equation}
    {{\nabla }_{{{\bm{\theta} }^{\mu }}}}J\approx {\nabla}_{\bm{a}}Q(\bm{s},\bm{a}){\nabla}_{{\bm{\theta}}^{\mu}}{\mu}(\bm{s}|{\bm{\theta}}^{\mu}).
\end{equation}

To obtain a more accurate estimate of the gradient of Q-values with respect to the policy parameters that is less noisy and more representative of the overall data, we take the mean of the sum of the gradients based on a mini-batch of data using Adam optimizer \cite{Adam} to update the policy parameters. The estimated gradient of Q-value by taking the mean of the sum can be expressed as:
\begin{equation}
\begin{aligned}
    &{{\nabla }_{{{\bm{\theta} }^{\mu }}}}J\approx \\
    &\frac{1}{B}\sum_{i}\bm{s}_{i}[{{{\nabla }_{\bm{a}}}Q(\bm{s},\bm{a}|{{\bm{\theta} }^{Q}}){{|}_{\bm{s}={\bm{s}_{i}},\bm{a}=\mu ({\bm{s}_{i}})}}{{\nabla }_{{{\bm{\theta} }^{\mu}}}}\mu (\bm{s}|{{\bm{\theta} }^{\mu}}){{|}_{{\bm{s}=\bm{s}_{i}}}}}].
\end{aligned}
\end{equation}

Following the optimization of the behaviour networks of the actor and the critic, we proceed to update their respective target networks. This update is performed using a soft update rule, which allows the target network to slowly track the parameters of the learned networks. The optimization process of the target networks can be expressed as: 
\begin{equation}    
    \bm{\theta}' = {\tau}{\bm{\theta}}+(1-\tau)\bm{\theta}',
\end{equation}
where $\bm{\theta}'$ represents the parameters of target networks, $\bm{\theta}$ denotes the parameters of behaviour networks, and $\tau$ is a hyper-parameter referred to as ``soft update rate". This parameter controls the gradual adjustments of target network parameters to align with those of the behaviour networks, preventing abrupt policy changes during training. $\tau$ is usually defined as a positive value that is much smaller than 1 for slow update, i.e., $\tau \ll 1$. Furthermore, the receiver is trained using supervised learning. In this process, batches of state-action pairs are randomly sampled from the buffer. The state-action pairs are used as input, with their corresponding outputs serving as labels. The receiver network is optimized by minimizing the binary cross-entropy loss. The training approach allows our proposed solution to optimize the transceivers without requiring prior knowledge of the channel model.

\subsection{Transmitter Training}
\begin{figure}[!]
  \centering
  \includegraphics[scale=0.35]{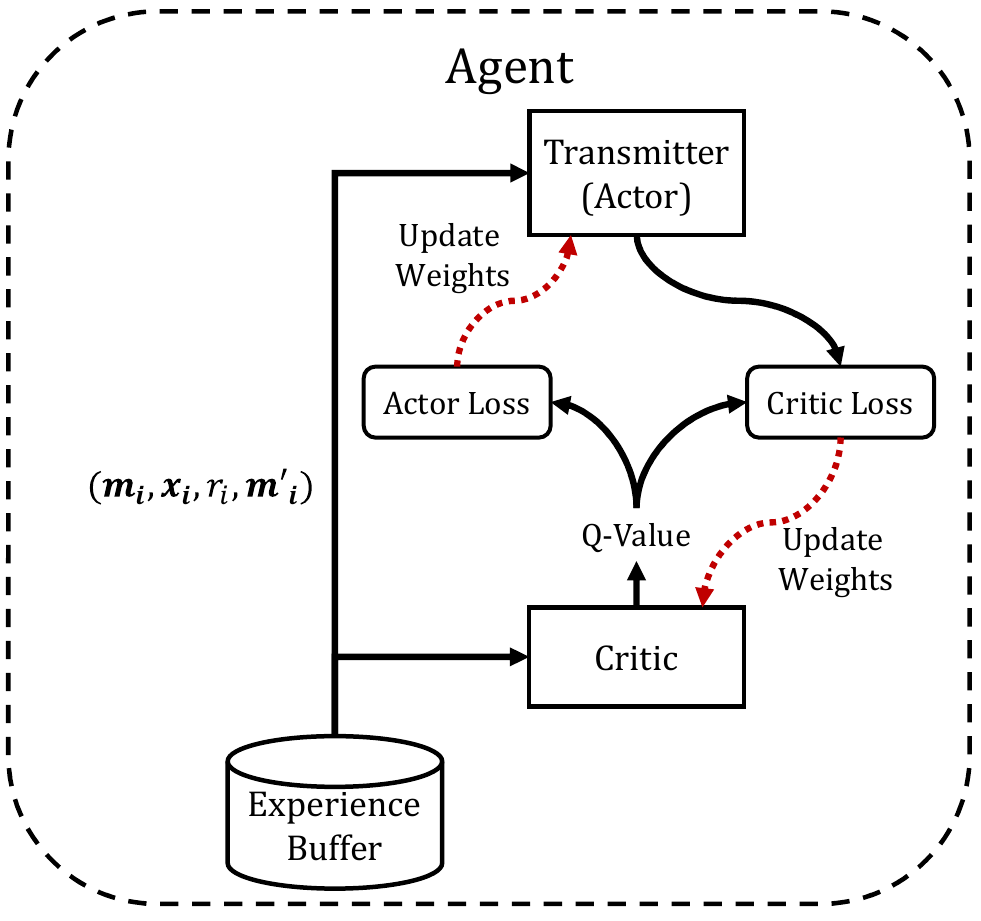}
  \caption{Training of transmitter.}
  \label{fig:TrainTransmitter}
\end{figure}

Fig.~\ref{fig:TrainTransmitter} demonstrates the training of the transmitter (actor) and the critic networks, where $\bm{m}_\mathrm{b}$, $\bm{x}_\mathrm{b}$ and ${r}_\mathrm{b}$ are the sampled batches of messages, encoded signals, and rewards, respectively. The sampled batches are used to calculate the actor loss and the critic loss, which are then used for the networks' optimizations. The actor calculates its loss by taking the negative expected value of the critic's Q-value for the sampled message and encoded signal batches pair, while the critic calculates its loss by taking the mean squared error between its estimated Q-value and the target Q-value obtained from the Bellman function. The actor network then updates its parameters using gradient ascent \cite{GradientAscent}, while the critic network updates its parameters using gradient descent \cite{GradientDescent}.

\begin{figure}[!]
  \centering
  \includegraphics[scale=0.32]{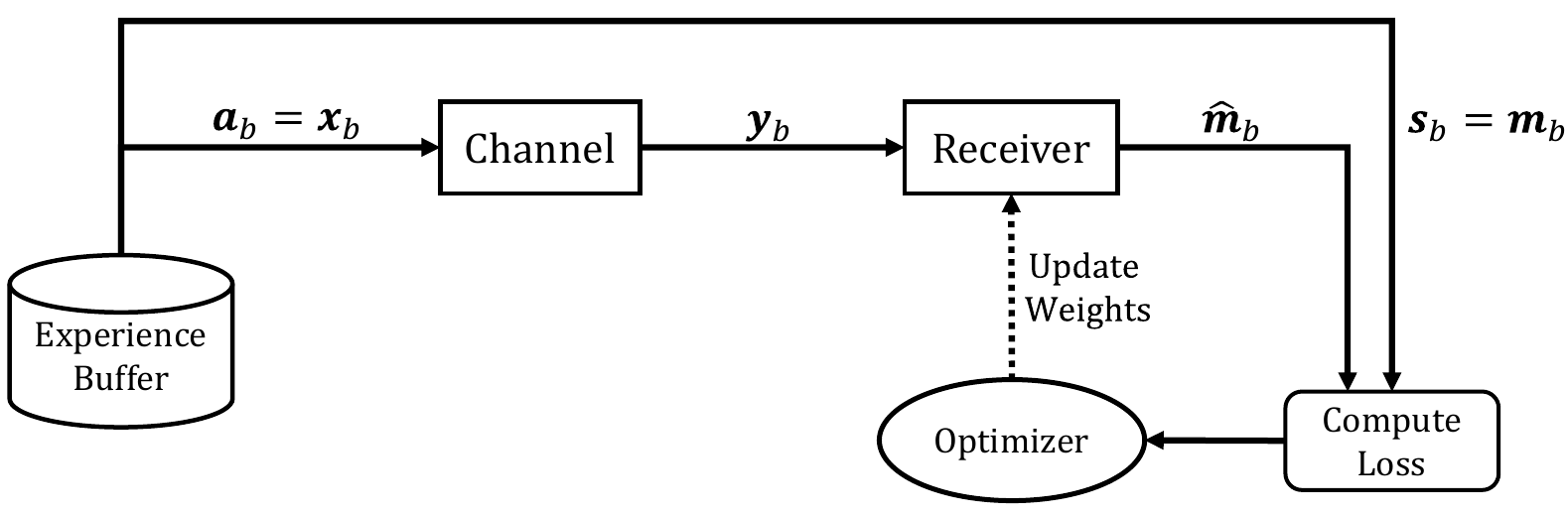}
  \caption{Training of receiver.}
  \label{fig:TrainReceiver} 
\end{figure}

\subsection{Receiver Training}

Fig. \ref{fig:TrainReceiver} shows the training process of the receiver model, where $\bm{x}_\mathrm{b}$ denotes the sampled encoded signal batch, $\bm{y}_\mathrm{b}$ denotes the received signal batch after channel distortion, $\bm{m}_\mathrm{b}$ denotes the sampled message batch and $\hat{\bm{m}}_\mathrm{b}$ denotes the estimated message batch generated by the receiver. At every time step, the encoded signal batch $\bm{x}_\mathrm{b}$ and its corresponding message batch $\bm{m}_\mathrm{b}$ are randomly sampled from the experience buffer. The received signal batch $\bm{y}_\mathrm{b}$ and the sampled message batch $\bm{m}_\mathrm{b}$ are considered as the input-output pair for training the receiver model in a supervised manner. The estimated message batch $\hat{\bm{m}}_\mathrm{b}$ is then used to calculate the binary cross-entropy loss, given the sampled message batch $\bm{m}_\mathrm{b}$ as the label. The calculated loss is used to update the weights of the receiver using the Adam optimizer~\cite{Adam}.

\section{Simulation Results}
\label{sec:evaluation}
\begin{figure*}[!]
	\centering
	\begin{subfigure}[b]{0.48\textwidth}
		\centering
		\includegraphics[width=\textwidth]{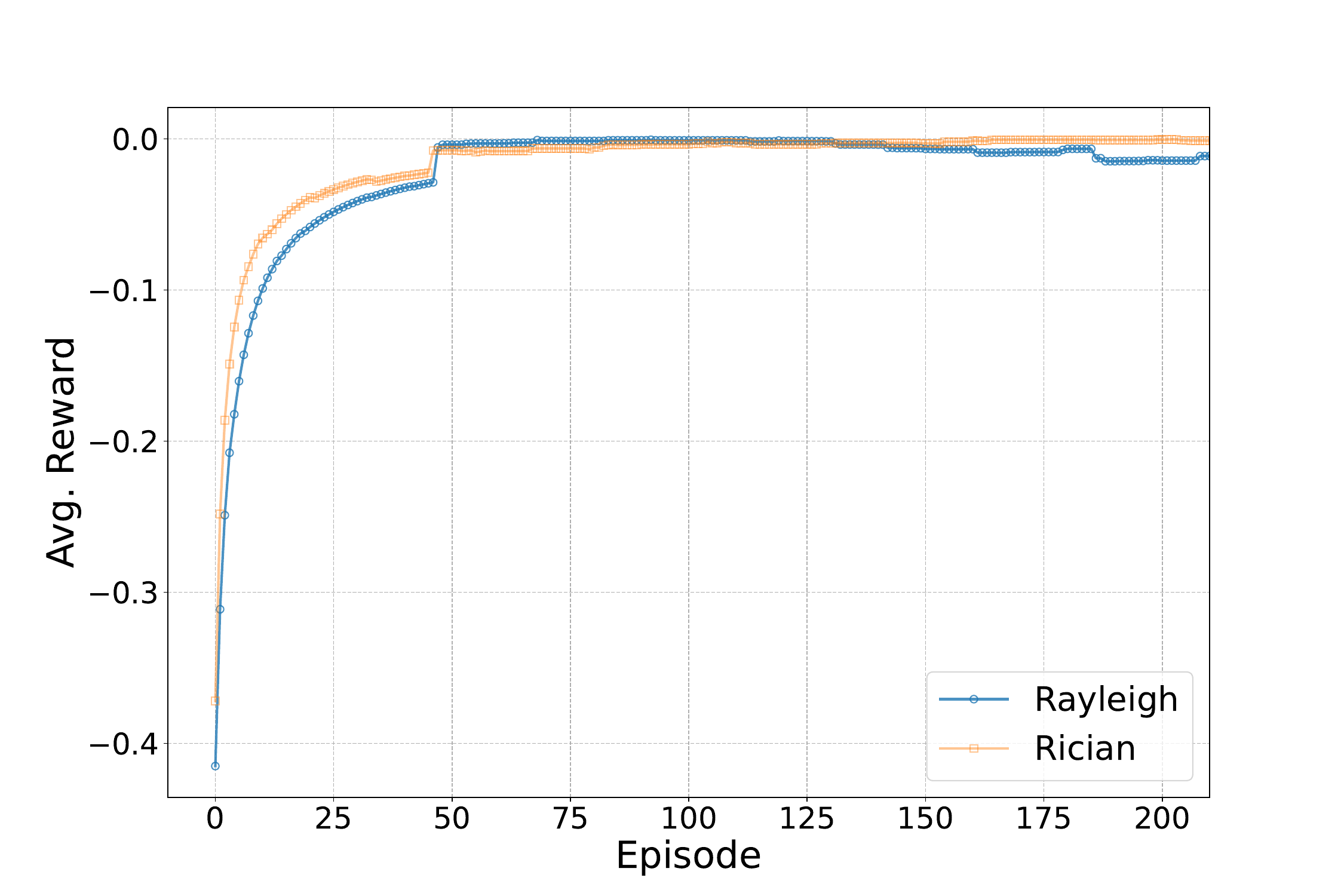}
		\caption{}
            \label{fig:reward_combined_a}
	\end{subfigure}
	\hspace{-0.9cm}
	\begin{subfigure}[b]{0.48\textwidth}
		\centering
		\includegraphics[width=\textwidth]{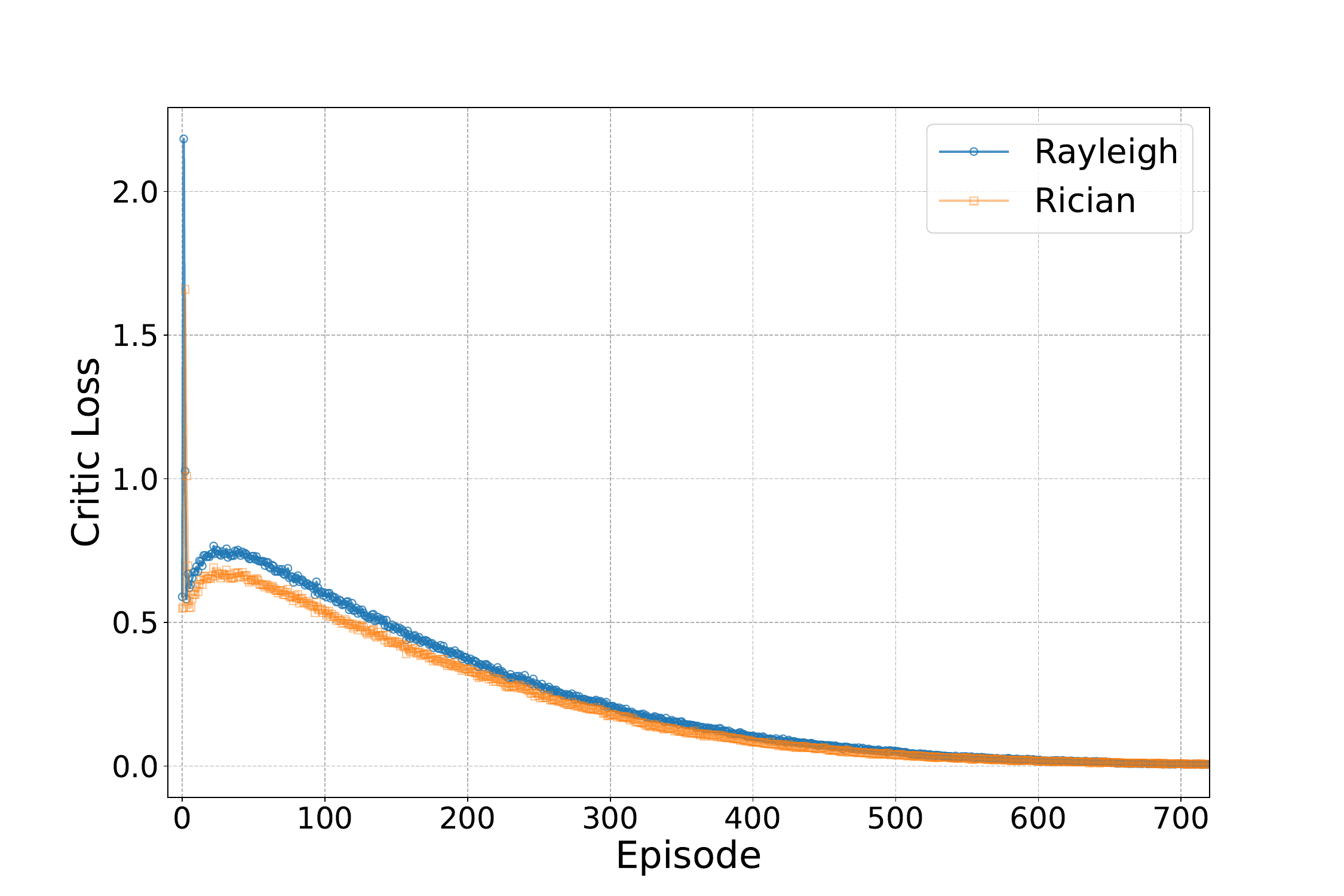}
		\caption{}
            \label{fig:reward_combined_b}
	\end{subfigure}
	\caption{(a) Averaged Episodic reward, and (b) Critic Loss Convergence with block length of 256 bits over fading channels.}
	\label{fig:reward_combined}
\end{figure*}

In this section, we will evaluate the performance of our proposed DDPG-based solution under Rician, Ryaleigh, and 3GPP channel models. In particular, the Rayleigh fading channel describes the wireless communication scenarios where there is no direct path but only none-line-of-sight (NLOS) paths during signal propagation. The amplitude of the randomly scattering signals can be described using the probability density function (p.d.f.) of Rayleigh distribution, as follows:
\begin{equation}
f(x) = \frac{x}{\sigma^2} e^{-\frac{x^2}{2\sigma^2}}, \quad x \geq 0,
\label{eq:rayleigh}
\end{equation}
where $x$ is the amplitude of the signal, $\sigma$ is the scale parameter that determines the spread of Rayleigh distribution, $2{\sigma^2}=\mathbb{E}(x^2)$ is the mean power of the signal, and $\mathbb{E}(.)$ denotes statistical averaging. The Rician fading channel, on the other hand, describes the wireless communication scenario where there are both line-of-sight (LOS) and NLOS components during signal propagation. The p.d.f. of Rician distribution reflects the combined influence of LOS and NLOS effects on the signal amplitude, which can be expressed as:
\begin{equation}
f(x) = \frac{2L}{\sigma^2} e^{-L-\frac{x^2}{2\sigma^2}} I_0\left(\frac{xL}{\sigma^2}\right), \quad x \geq 0,
\label{eq:rician}
\end{equation}
where $x$ is the amplitude of the signal, $\sigma$ is the scale parameter that determines the spread of Rician distribution, $L$ is the ratio of the power of the dominant path to the scattered path, $I_0$ is the modified Bessel function of the first kind.

The baselines for comparisons are the alternating training (AT) scheme proposed in~\cite{without}, the conditional GAN scheme proposed in~\cite{GAN} and the residual-aided GAN-based scheme proposed in~\cite{ra-gan}. The classic baseline considers LDPC channel coding for large block sizes, specifically 128 and 256 bits. Since LDPC requires a block size larger than 24 bits, it is only applied to these larger block sizes. The supervised learning baseline assumes that the channel gradients are known to the transmitter NN, with the transmitter and receiver together forming a complete autoencoder, referred to as AE. The learning rates of both the transmitter and the receiver are set as 0.001 for all the baselines. The following simulations are evaluated by BLER over different signal-to-noise ratios (SNRs), ranging from 0 dB to 20 dB. 


\subsection{Parameter Setting}
For the hyper-parameters settings, the soft update rate $\tau$ is set to 0.005 for slow updates of the target networks. The learning rates of the actor and critic networks are set as 0.002 and 0.001, respectively. Training consists of 30,000 episodes with 500 time steps per episode. The discount factor $\gamma$ is set at 0.01. The DDPG-based solution is trained on block fading channels with different block sizes, i.e., 8, 128, and 256 information bits. The Rician factor is 1, indicating that the portions of the LOS and NLOS components are equivalent.

\subsection{Performance Evaluation}

\begin{figure*}[!]
	\centering
	\begin{subfigure}[b]{0.48\textwidth}
		\centering
		\includegraphics[width=\textwidth]{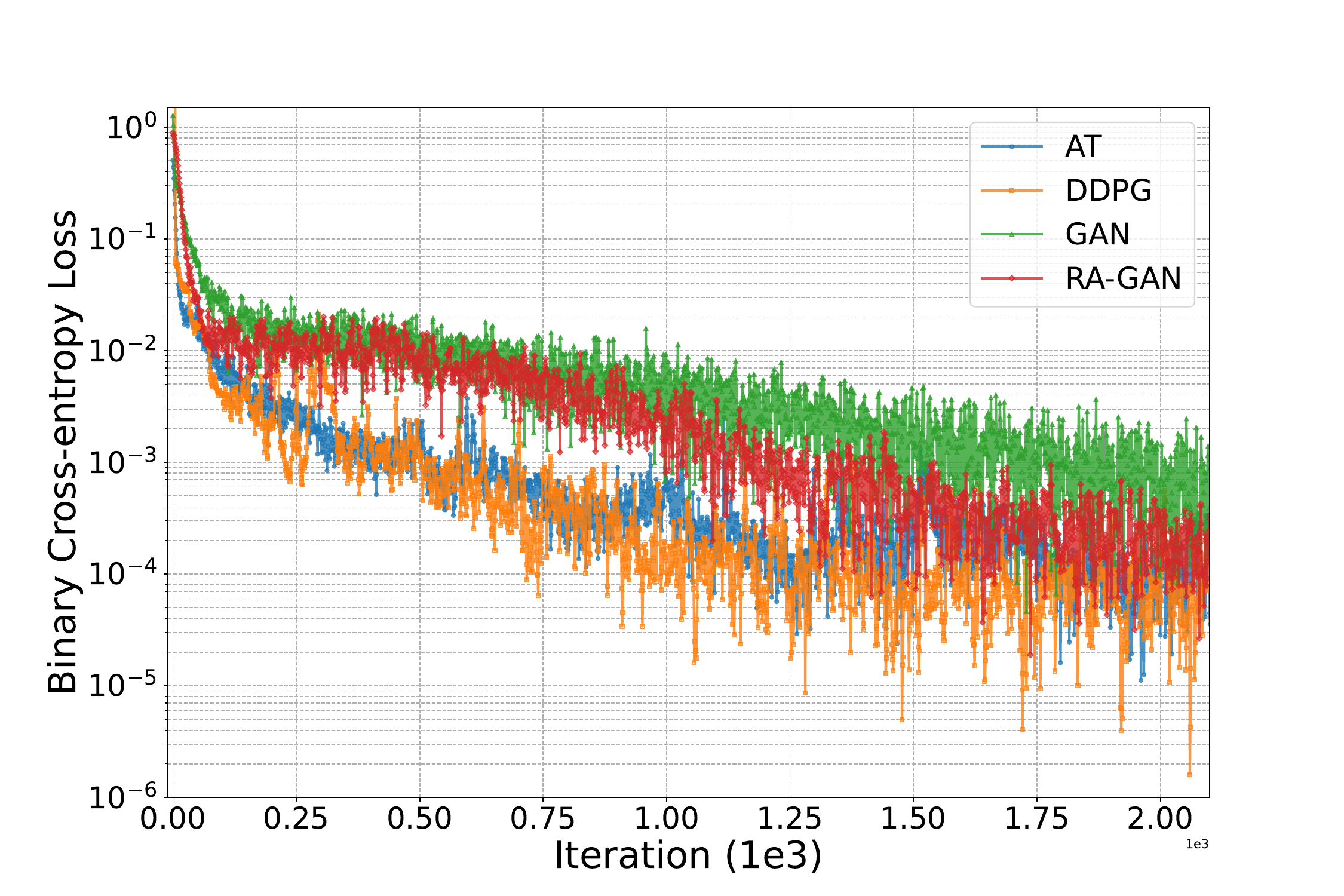}
		\caption{}
            \label{fig:Convergence_a}
	\end{subfigure}
	\hspace{-0.9cm}
	\begin{subfigure}[b]{0.48\textwidth}
		\centering
		\includegraphics[width=\textwidth]{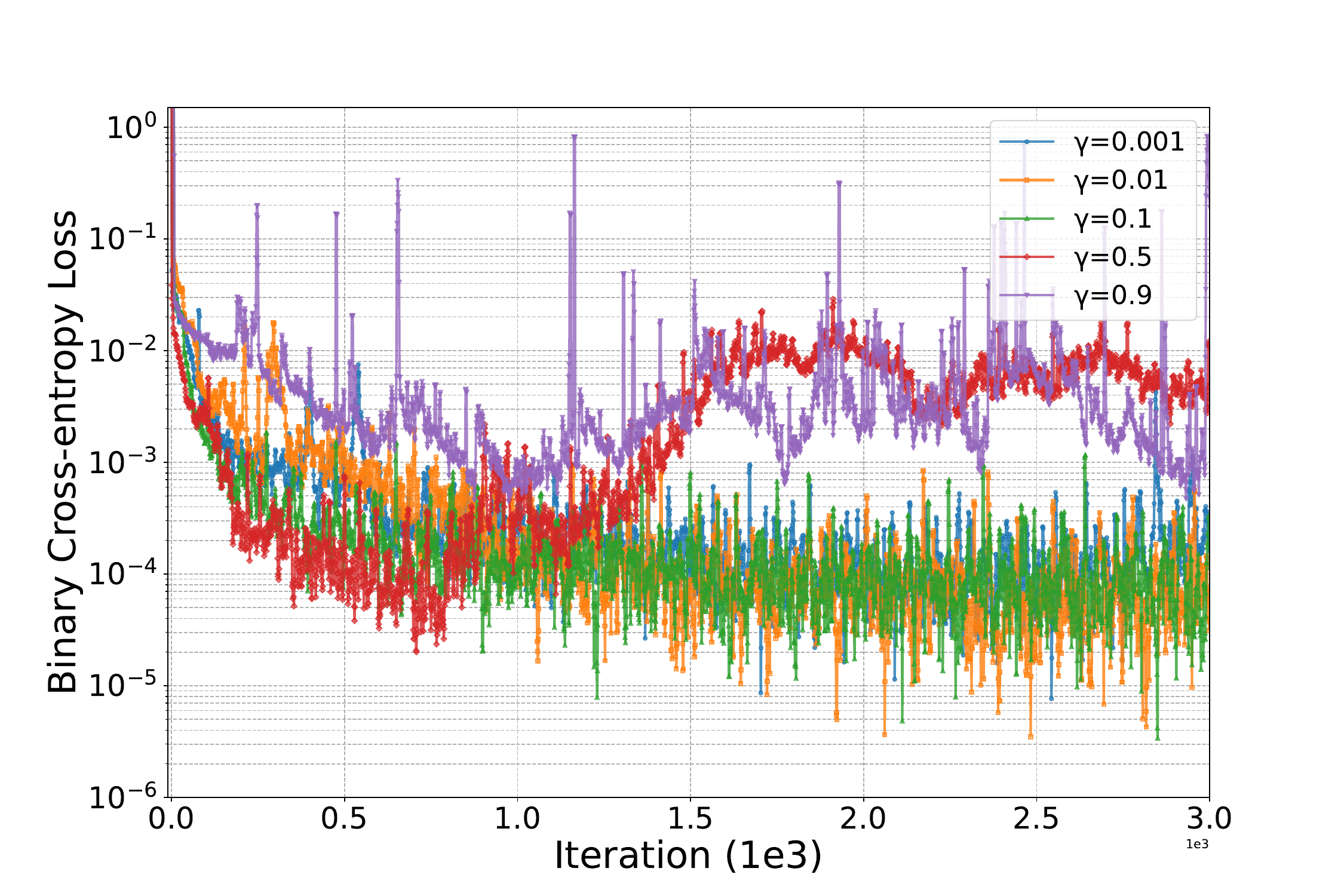}
		\caption{}
            \label{fig:Convergence_b}
	\end{subfigure}
	\caption{(a) Receiver Loss Convergence for different schemes, and (b) Receiver Loss Convergence for different discount factors.}
	\label{fig:Convergence}
\end{figure*}



\begin{figure*}[!]
	\centering
	\begin{subfigure}[b]{0.35\textwidth}
		\centering
		\includegraphics[width=\textwidth]{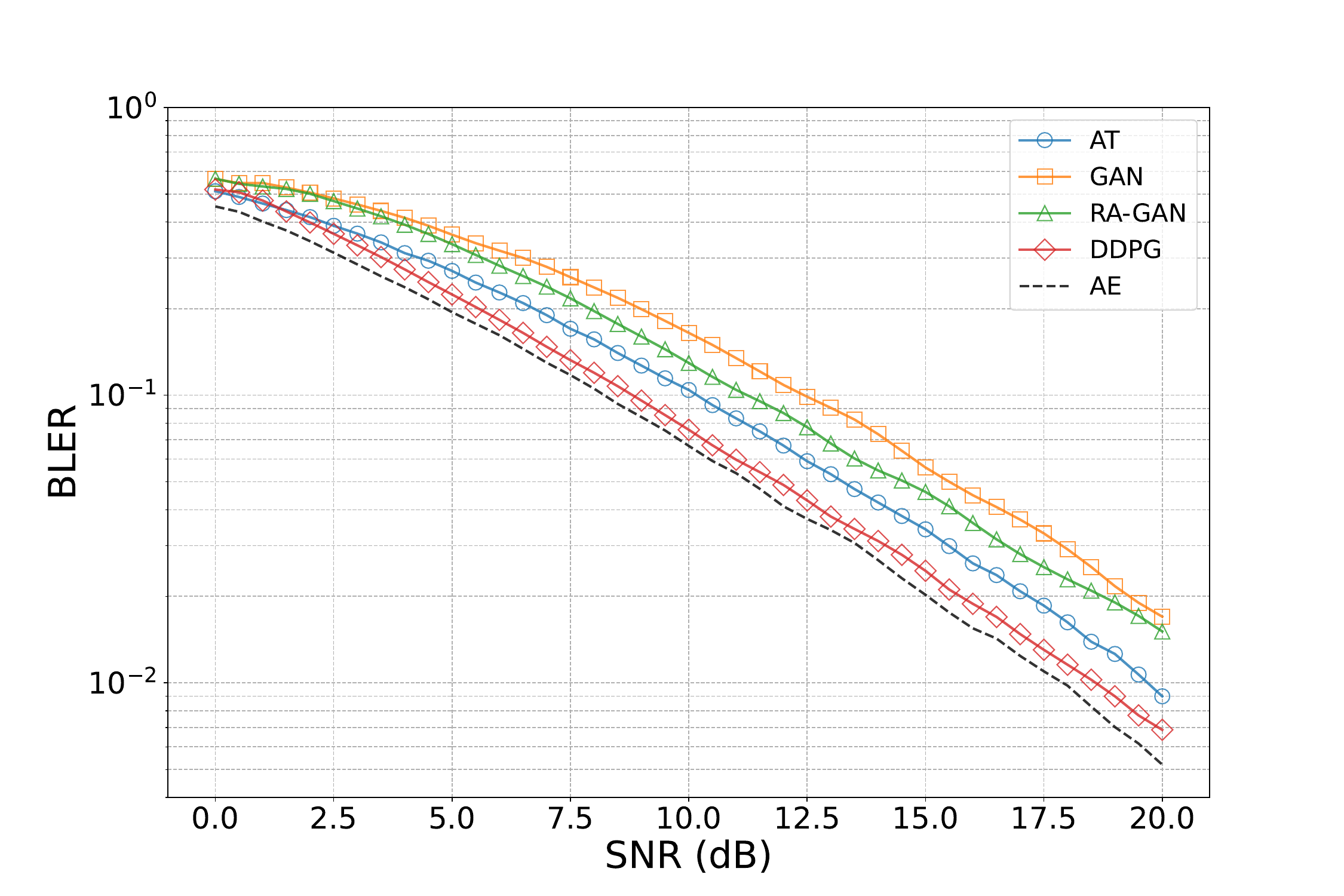}
		\caption{}
            \label{fig:rayleigh_20_8}
	\end{subfigure}
	\begin{subfigure}[b]{0.35\textwidth}
		\centering
		\includegraphics[width=\textwidth]{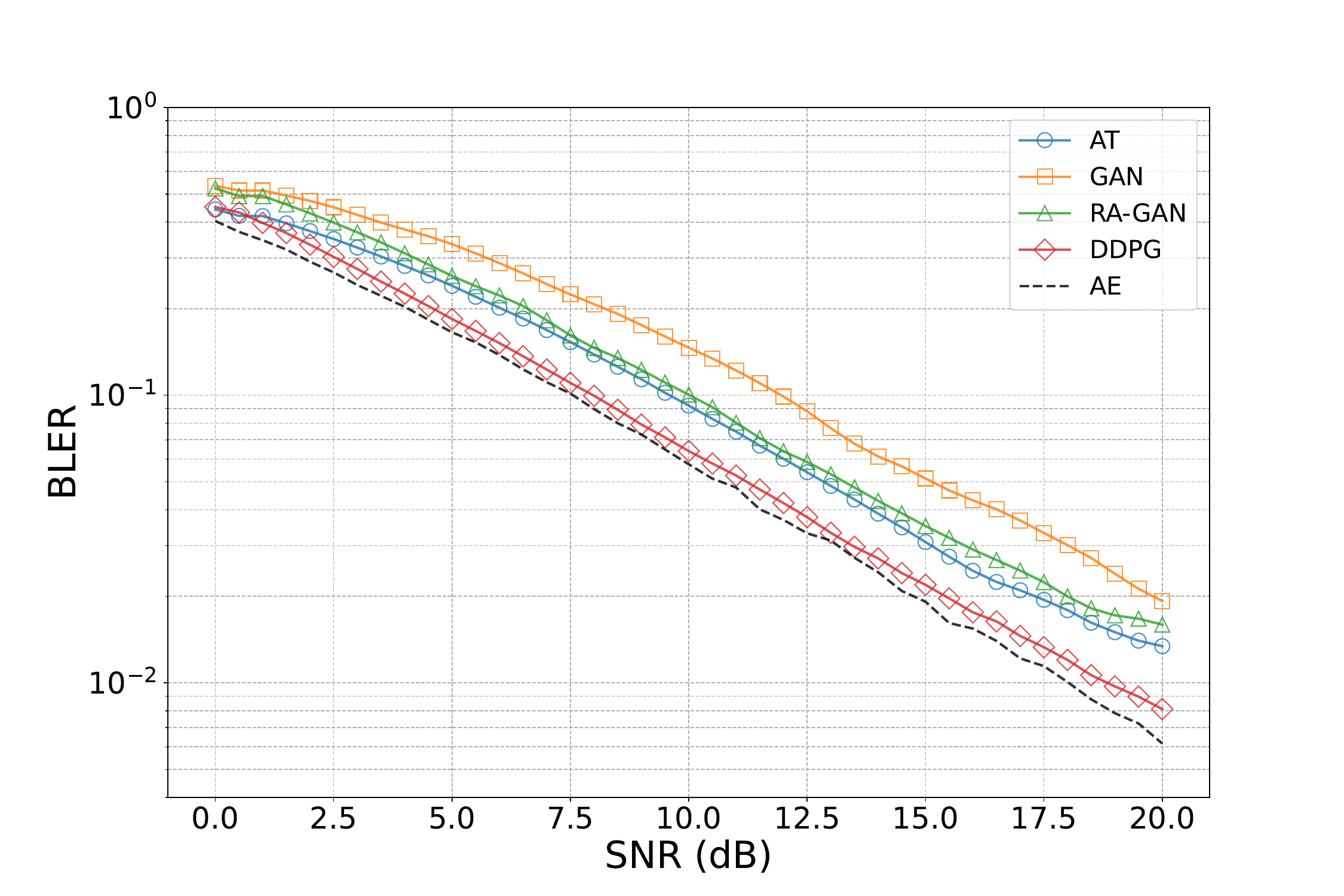}
		\caption{}
            \label{fig:rayleigh_10_8}
	\end{subfigure}
        
        \begin{subfigure}[b]{0.35\textwidth}
		\centering
		\includegraphics[width=\textwidth]{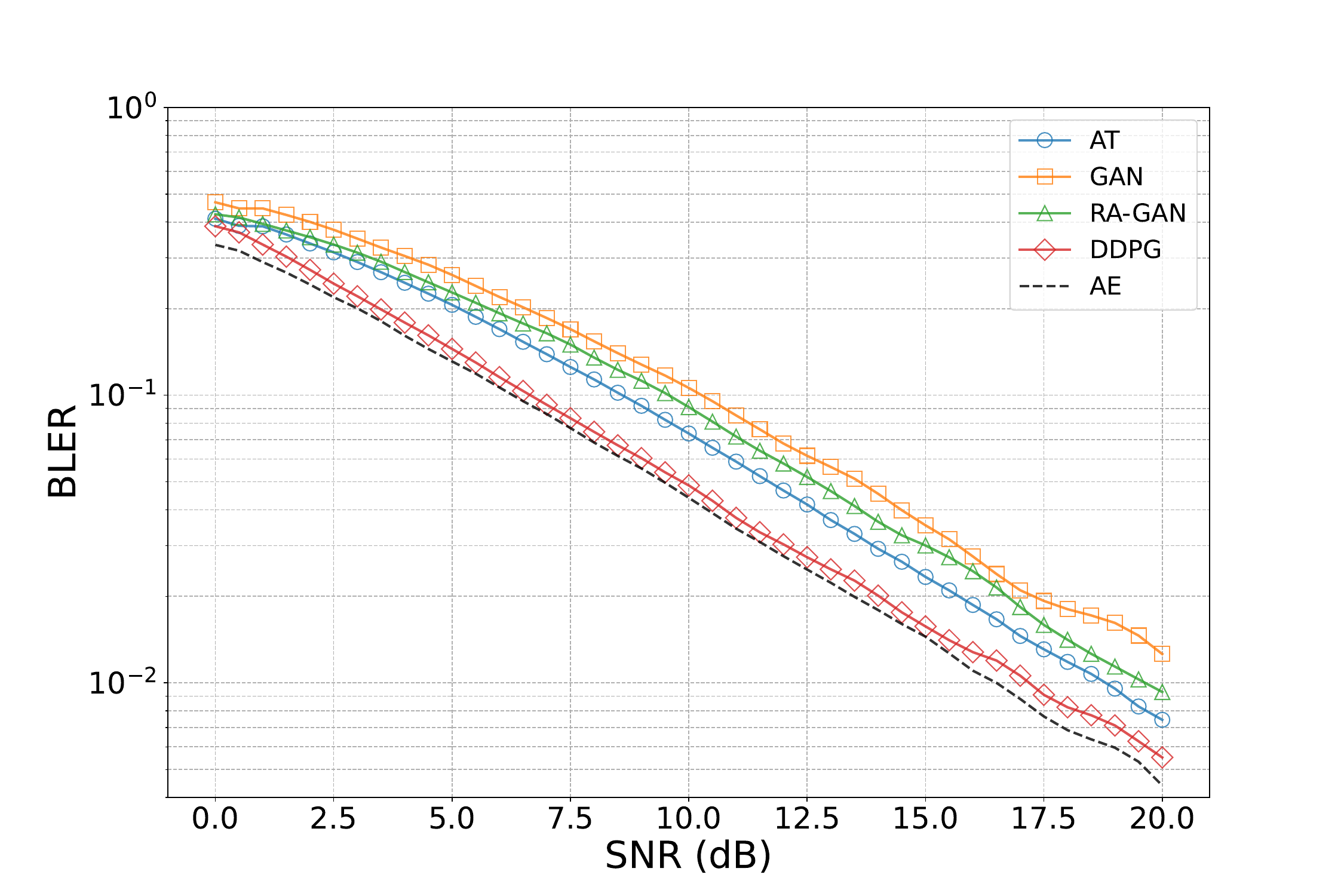}
		\caption{}
            \label{fig:rician_20_8}
	\end{subfigure}
	\begin{subfigure}[b]{0.35\textwidth}
		\centering
		\includegraphics[width=\textwidth]{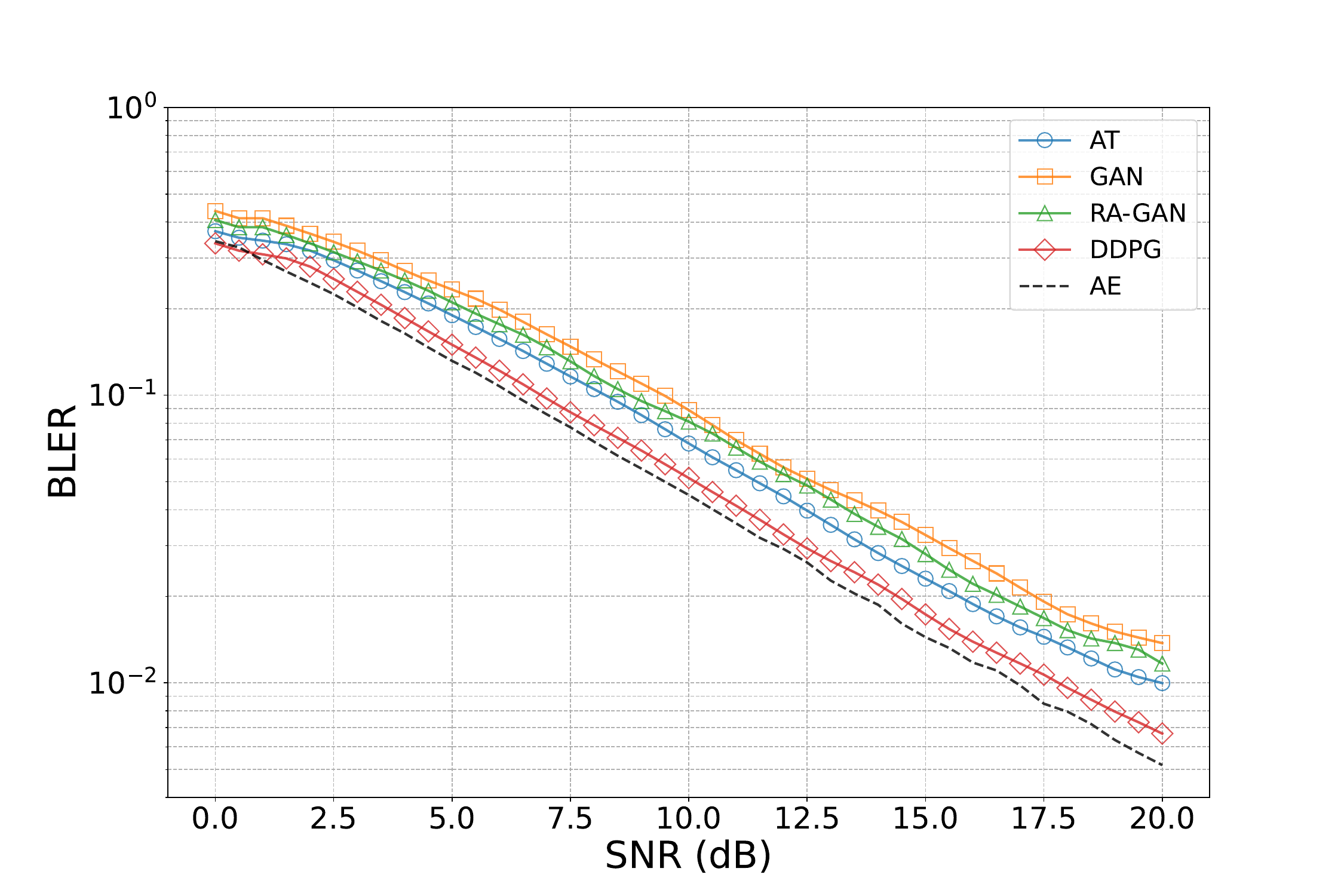}
		\caption{}
            \label{fig:rician_10_8}
	\end{subfigure}
	\caption{BLER vs. SNR with block size of 8 trained at (a) Rayleigh 20dB, (b) Rayleigh 10dB, (c) Rician 20dB, (d) Rician 10dB.}
	\label{fig:Rayleigh8}
\end{figure*}

\begin{figure*}[!]
	\centering
	\begin{subfigure}[b]{0.35\textwidth}
		\centering
		\includegraphics[width=\textwidth]{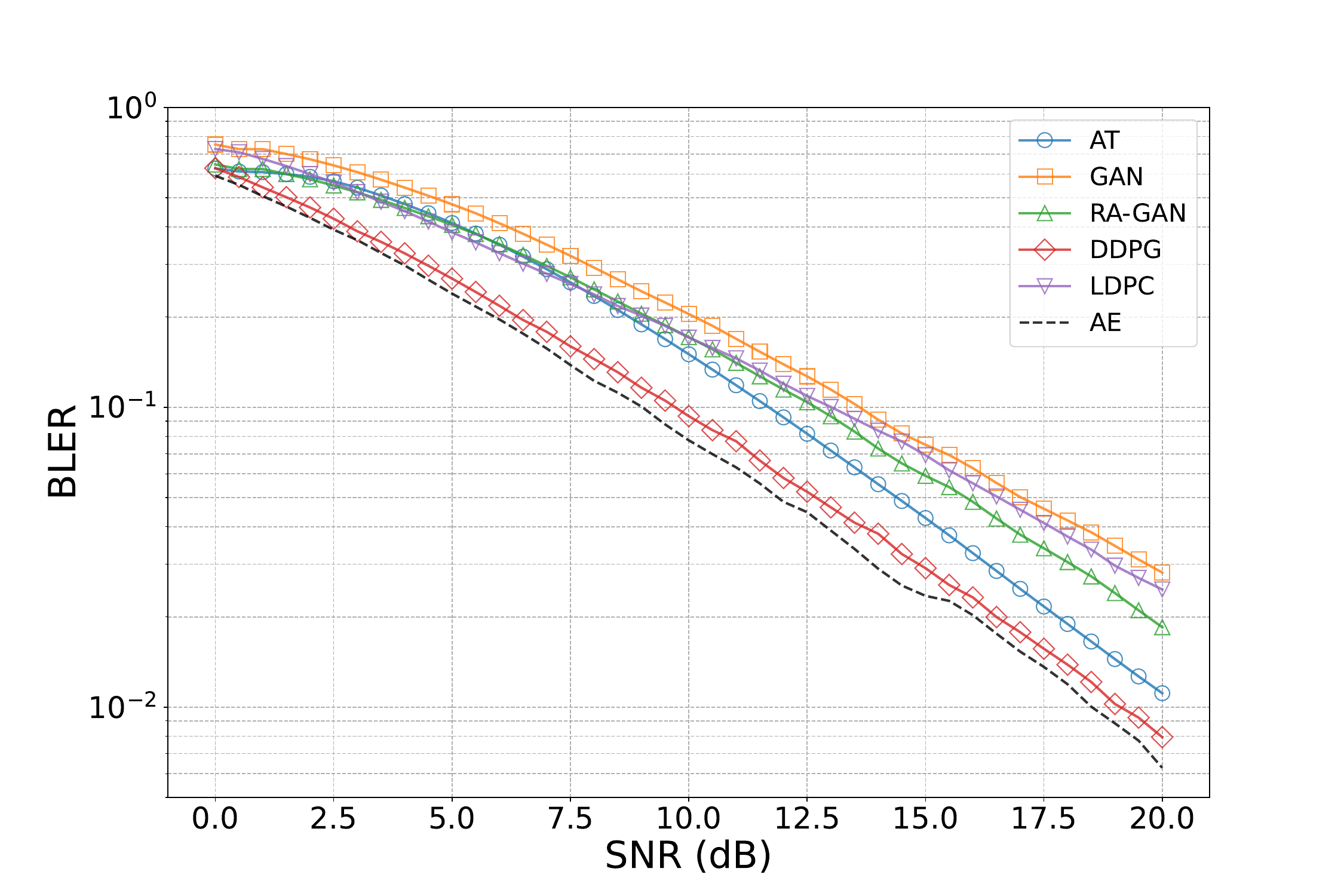}
		\caption{}
            \label{fig:rayleigh_20_128}
	\end{subfigure}
	\begin{subfigure}[b]{0.35\textwidth}
		\centering
		\includegraphics[width=\textwidth]{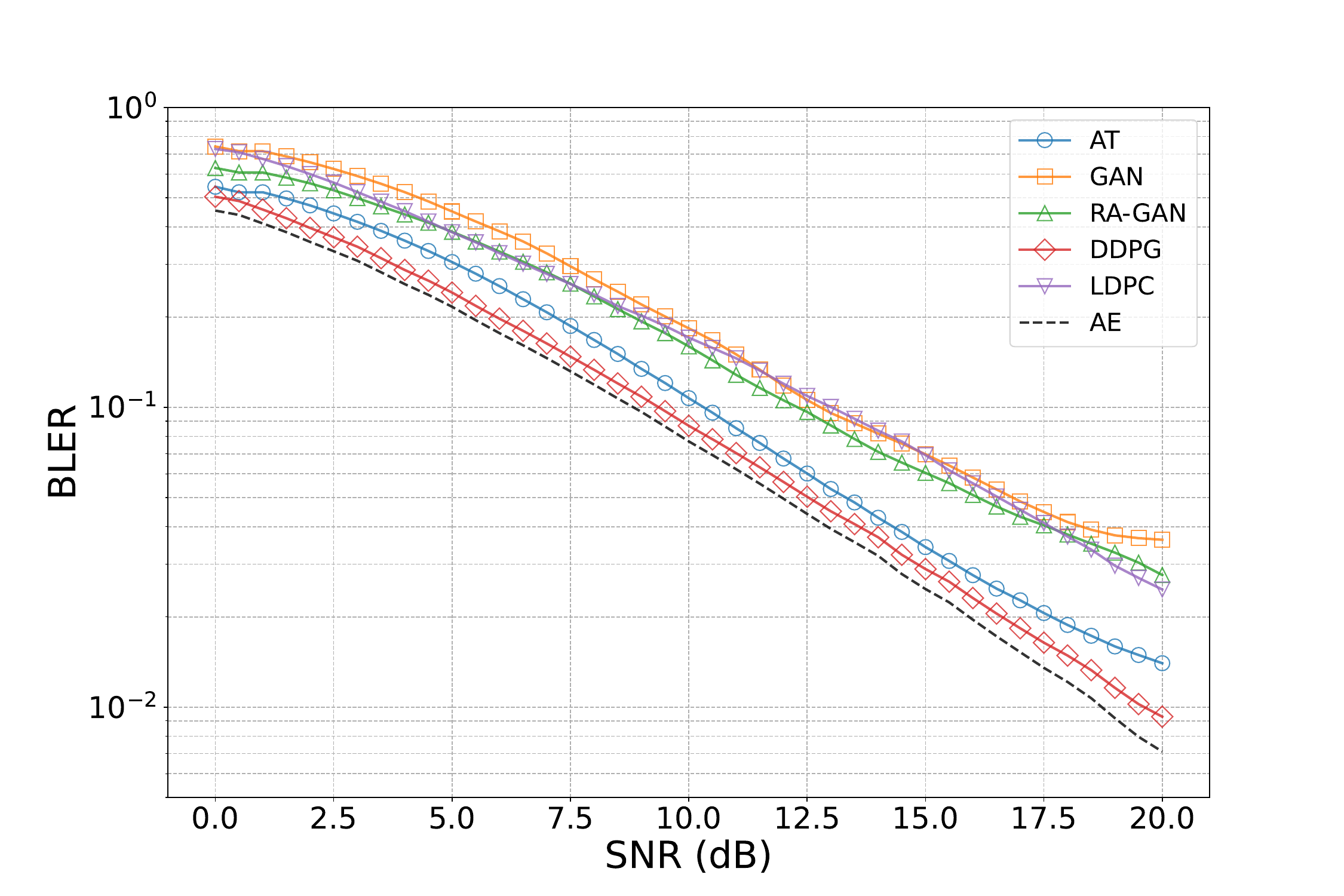}
		\caption{}
            \label{fig:rayleigh_10_128}
	\end{subfigure}
        
        \begin{subfigure}[b]{0.35\textwidth}
		\centering
		\includegraphics[width=\textwidth]{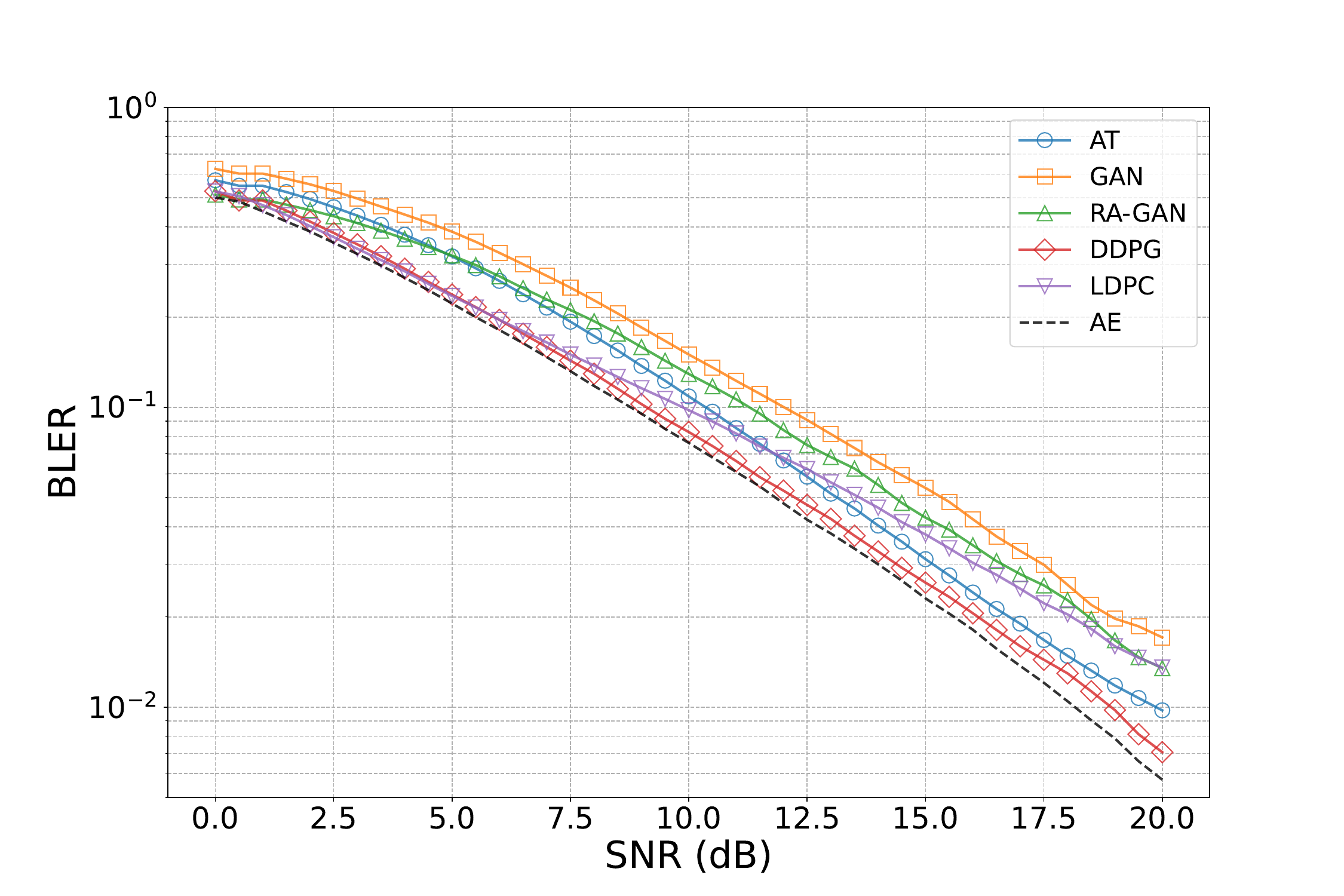}
		\caption{}
            \label{fig:rician_20_128}
	\end{subfigure}
	\begin{subfigure}[b]{0.35\textwidth}
		\centering
		\includegraphics[width=\textwidth]{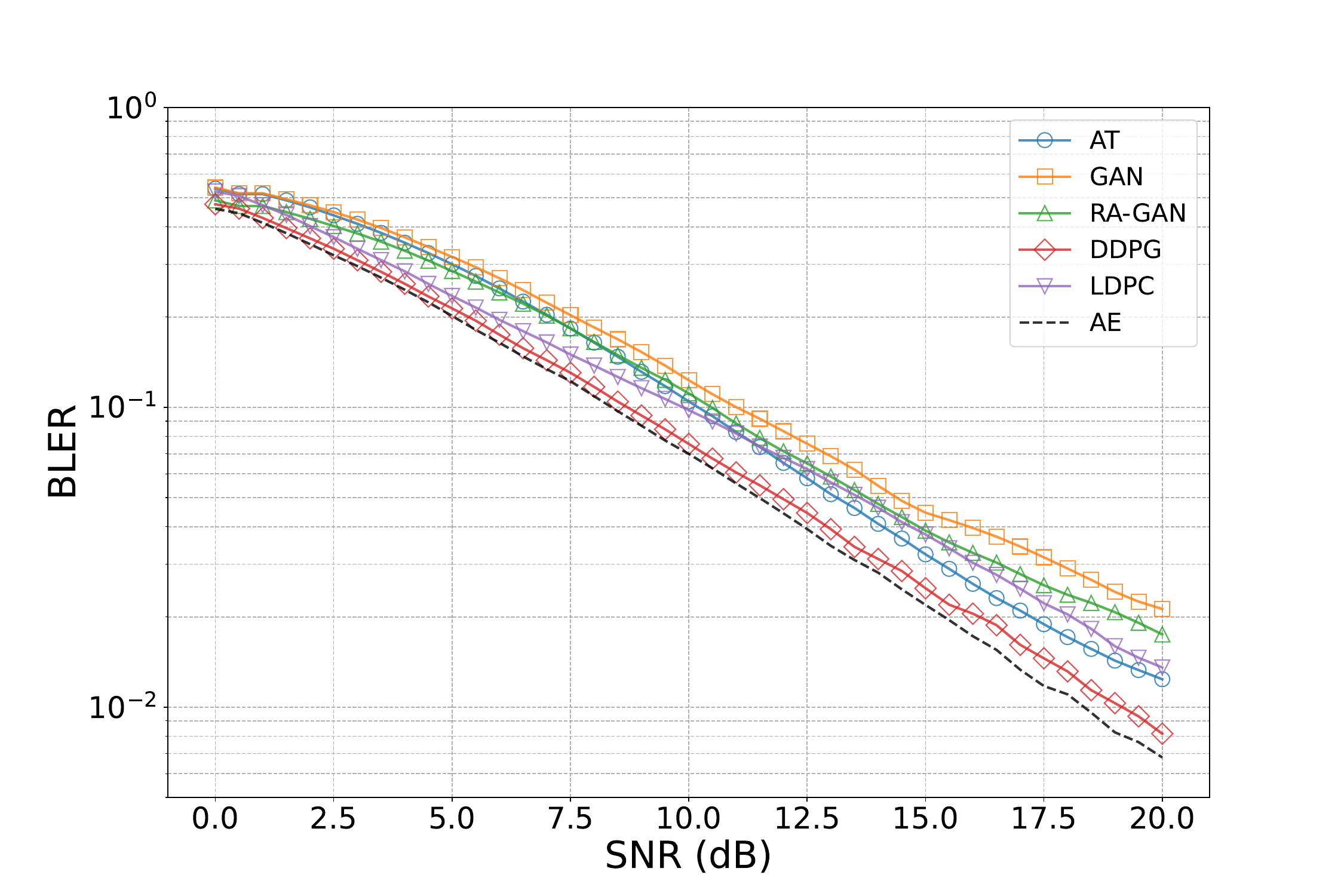}
		\caption{}
            \label{fig:rician_10_128}
	\end{subfigure}
	\caption{BLER vs. SNR with block size of 128 trained at (a) Rayleigh 20dB, (b) Rayleigh 10dB, (c) Rician 20dB, (d) Rician 10dB.}
	\label{fig:Rayleigh128}
\end{figure*}

\begin{figure*}[!]
	\centering
	\begin{subfigure}[b]{0.35\textwidth}
		\centering
		\includegraphics[width=\textwidth]{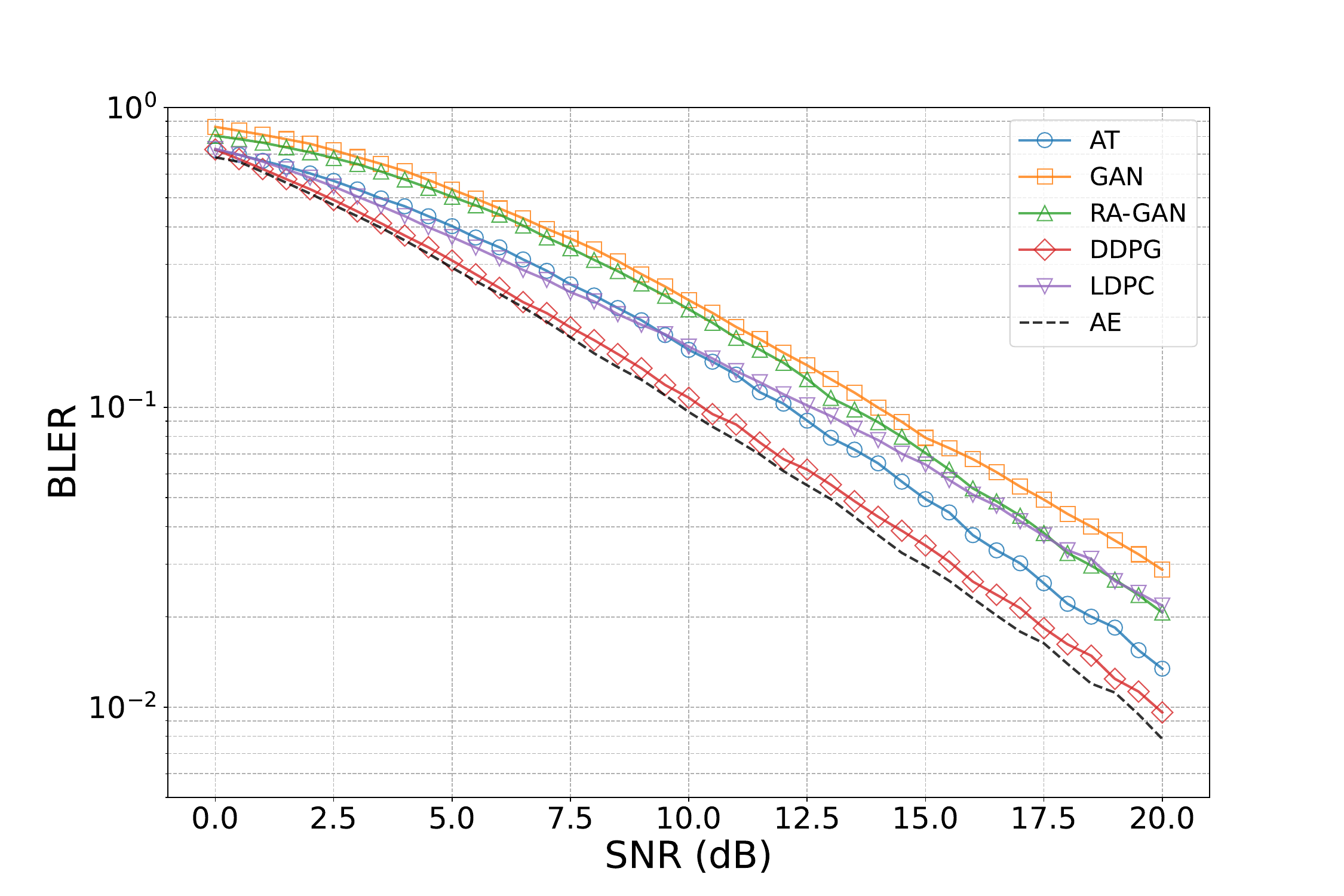}
		\caption{}
            \label{fig:rayleigh_20_256}
	\end{subfigure}
	\begin{subfigure}[b]{0.35\textwidth}
		\centering
		\includegraphics[width=\textwidth]{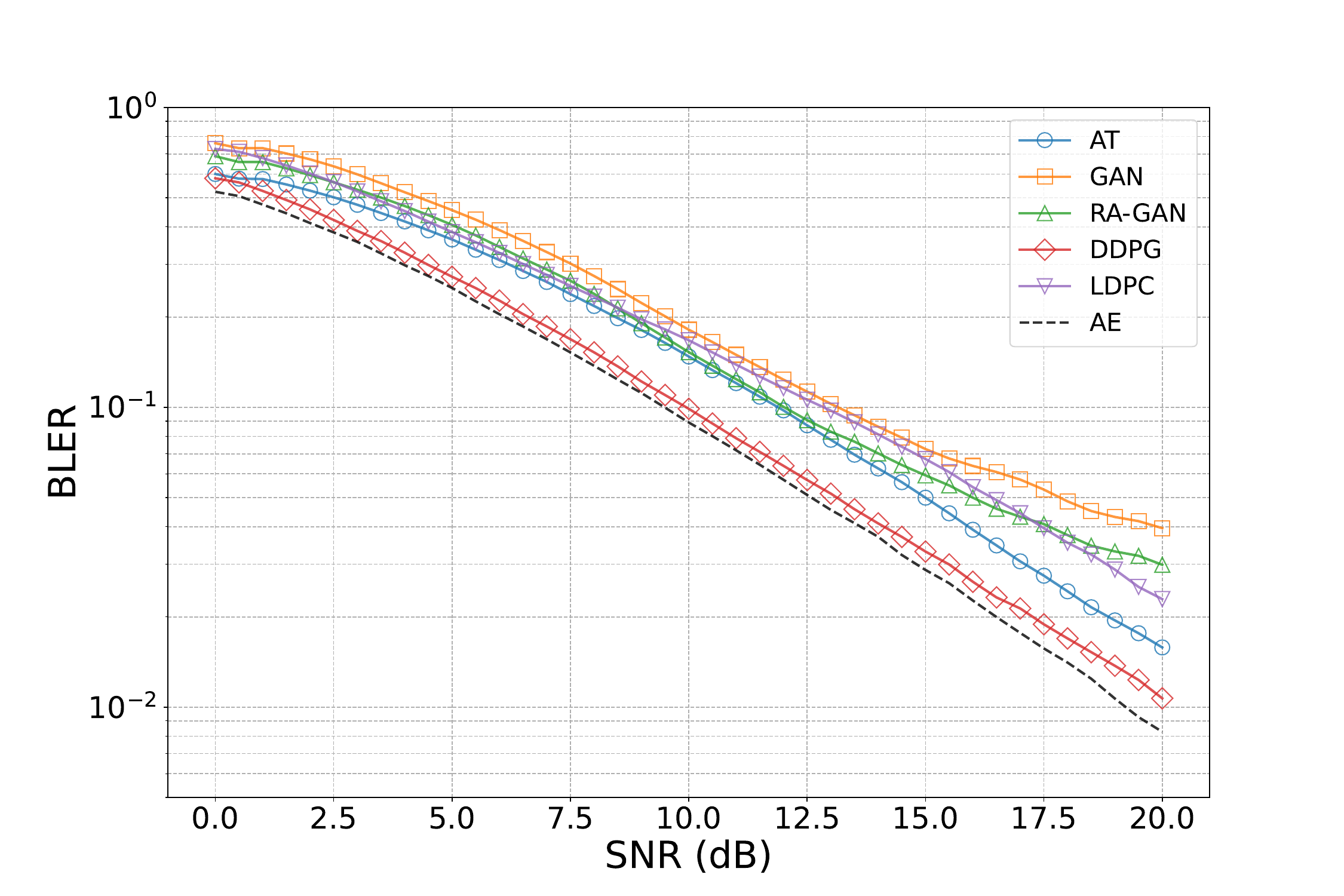}
		\caption{}
            \label{fig:rayleigh_10_256}
	\end{subfigure}
        
        \begin{subfigure}[b]{0.35\textwidth}
		\centering
		\includegraphics[width=\textwidth]{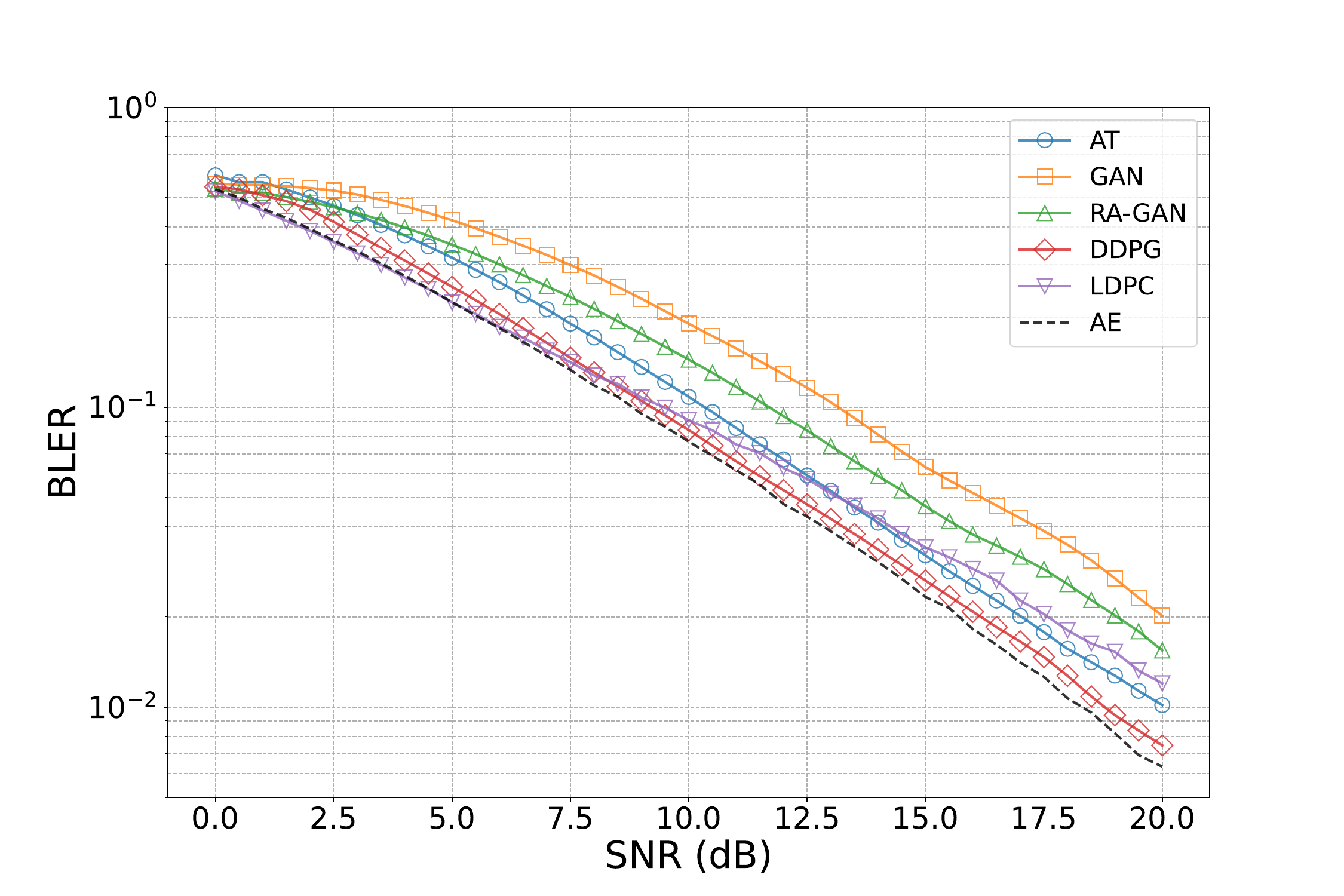}
		\caption{}
            \label{fig:rician_20_256}
	\end{subfigure}
	\begin{subfigure}[b]{0.35\textwidth}
		\centering
		\includegraphics[width=\textwidth]{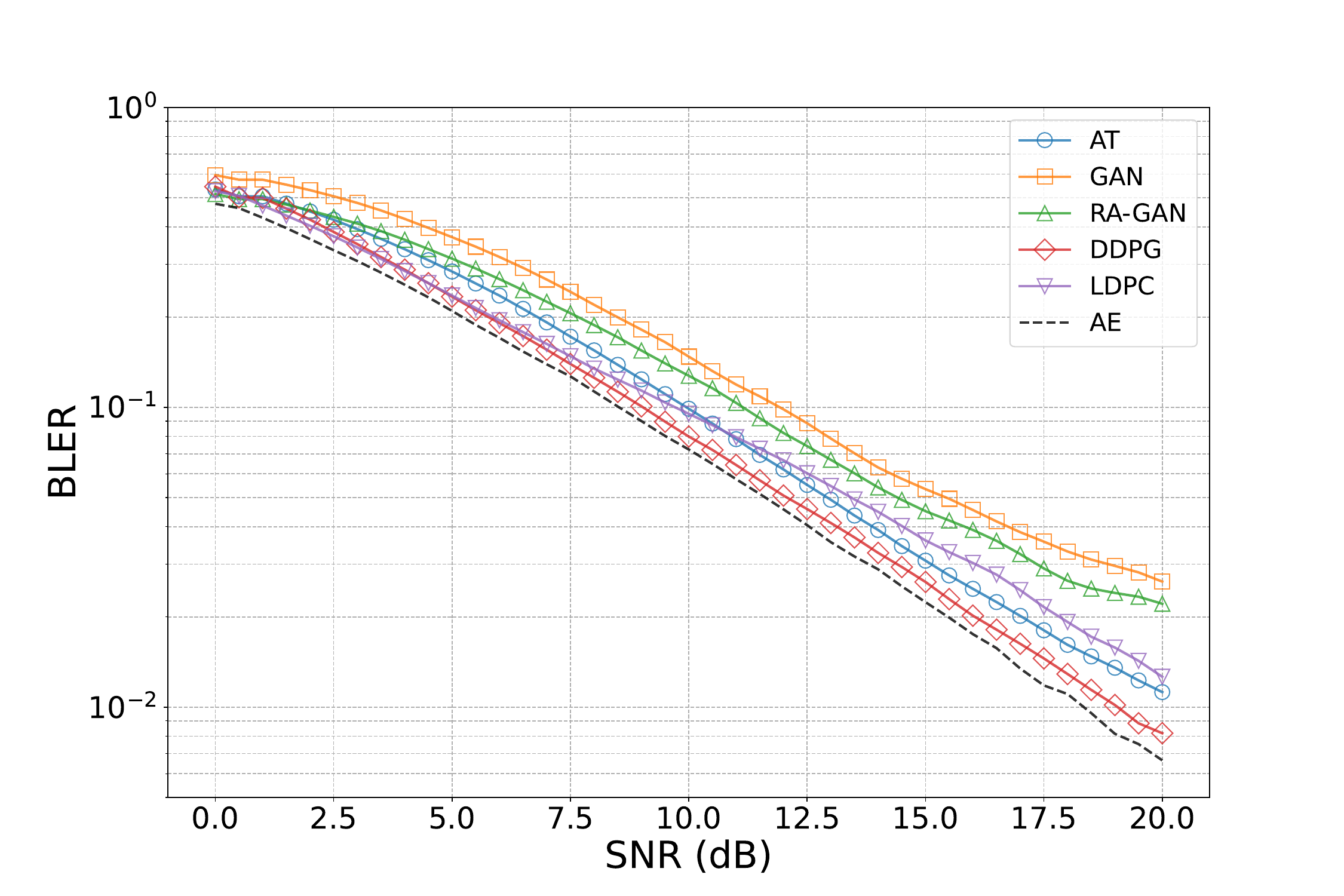}
		\caption{}
            \label{fig:rician_10_256}
	\end{subfigure}
	\caption{BLER vs. SNR with block size of 256 trained at (a) Rayleigh 20dB, (b) Rayleigh 10dB, (c) Rician 20dB, (d) Rician 10dB.}
	\label{fig:Rayleigh256}
\end{figure*}

\begin{figure}[!]
\centering
    \centering
    \includegraphics[width=0.48\textwidth]{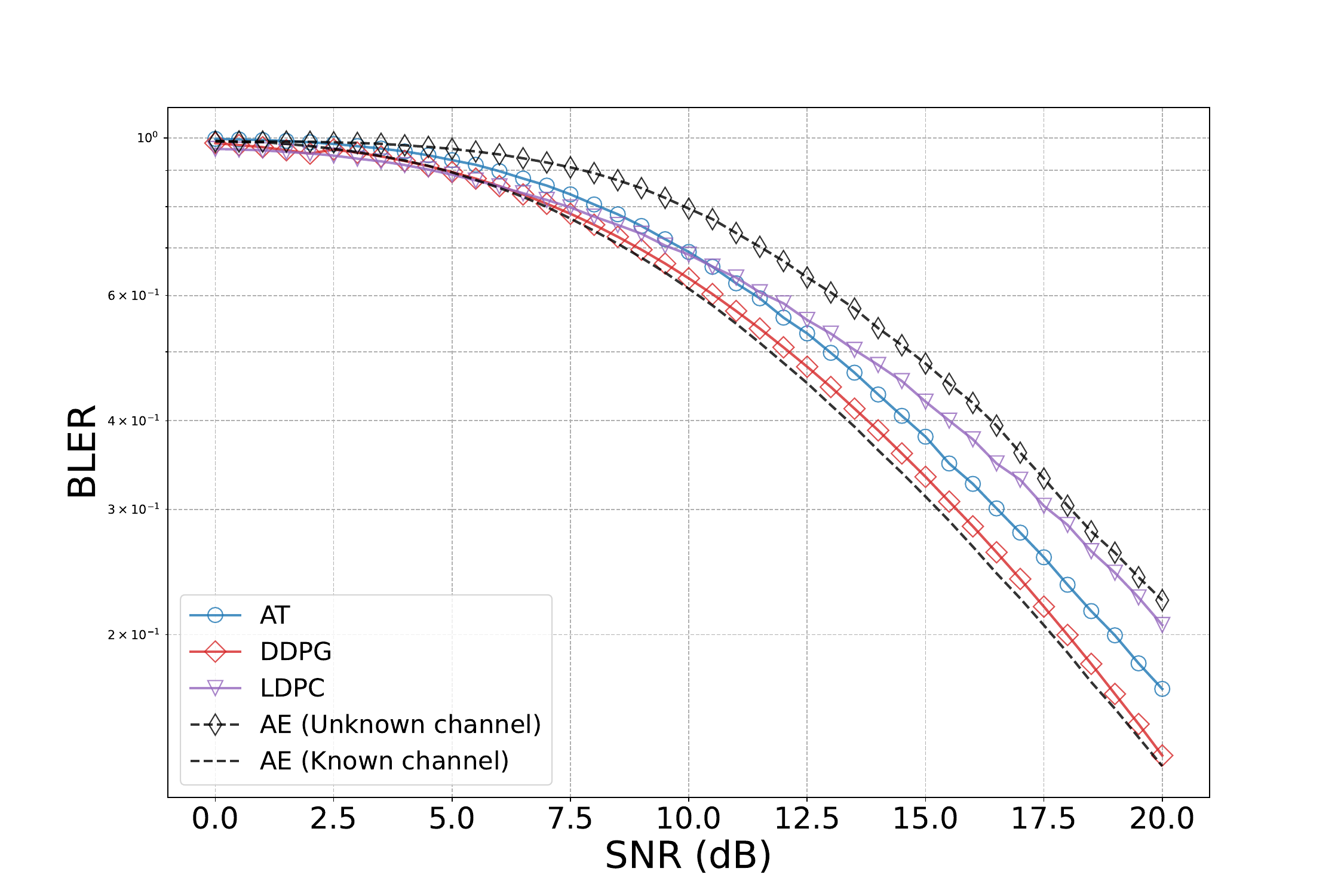}
\caption{BLER vs. SNR on 3GPP.}
    \label{fig:3gpp}
\end{figure}

\subsubsection{Training Process}

We first evaluate the training process of the proposed DDPG approach trained at 20dB with 256-bit block length over both Rayleigh and Rician fading channels, as shown in Fig.~\ref{fig:reward_combined}. As can be seen in Fig.~\ref{fig:reward_combined_a}, the averaged episodic rewards can converge close to zero within 200 episodes. It is worth noting that the reward sent from the receiver to the transmitter is the negative value of the receiver's training loss, as shown in (\ref{eq:rewardfunction}). As such, an averaged episodic reward approaching zero indicates a high detection accuracy at the receiver. This verifies the effectiveness of our proposed DDPG-based solution for training E2E communication systems without prior channel knowledge. In Fig.~\ref{fig:reward_combined_b}, the critic loss converges within 700 episodes, indicating that the critic network effectively learns to minimize the difference between the target Q-value and the estimated critic value. This convergence enables the network to provide more accurate guidance to the transmitter, leading to better optimization of the system. In the following, we evaluate the detection performance of our proposed solution under different settings in terms of BLER and convergence rate.

\subsubsection{Convergence Rate}

We conduct a convergence analysis by comparing the convergence rates of the receiver loss between the proposed DDPG-based solution and the baseline schemes, using a block size of 256 bits over a Rayleigh fading channel, as shown in Fig.~\ref{fig:Convergence_a}. Clearly, the proposed DDPG-based scheme exhibits faster convergence compared to the baseline schemes. Specifically, the proposed DDPG scheme converges to approximately $3 \times 10^{-5}$ for the receiver loss within 2,000 iterations, whereas the baseline methods do not achieve such low values. This demonstrates the effectiveness of the proposed DDPG-based scheme in rapidly adapting to new channel scenarios with significantly shorter training times compared to state-of-the-art solutions. 

Next, we evaluate the loss convergence of the proposed DDPG-based solution across different discount factors $\gamma$, specifically evaluating values of 0.001, 0.01, 0.1, 0.5 and 0.9, as illustrated in Fig. \ref{fig:Convergence_b}. The results demonstrate that an increase in discount factor negatively impacts loss convergence. Notably, high discount factors, such as $\gamma=0.5$ and $\gamma=0.9$, exhibit significant divergence. In contrast, smaller discount factors, including 0.001, 0.01, and 0.1, can reach a similar steady state after 1,000 iterations. This behaviour suggests that a small discount factor better aligns with the multi-armed bandit scenario, where the adjacent transmitted signals are independent and uncorrelated.

\subsubsection{Small Block Size}
First, we consider the cases when transmitting 8 information bits per message and observe the performance of our proposed DDPG-based solution. In particular, in Fig.~\ref{fig:rayleigh_20_8}, our proposed solution can obtain superior performance compared to other baselines. This demonstrates that the actor and critic networks in our proposed DDPG-based approach can efficiently learn the training process of the receiver and channel conditions through the feedback reward. Moreover, Fig.~\ref{fig:rayleigh_10_8} suggests that training the proposed model at an SNR of 10 dB results in a flatter performance curve compared to training at 20 dB. This leads to better performance in the lower SNR range but comes at the cost of reduced performance in the high SNR regime.

We then consider the Rician fading channel and observe the BLER performance of our proposed solution, as shown in Fig.~\ref{fig:rician_20_8} and Fig.~\ref{fig:rician_10_8}. Similar to Rayleigh channel scenarios, our proposed approach can also achieve better BLER performance than all the baseline schemes. This demonstrates the effectiveness of our solution in different wireless settings. 

\subsubsection{Large Block Size}

Next, we compare our proposed solution with the AT scheme~\cite{without}, the GAN solution~\cite{GAN}, the RA-GAN solution~\cite{ra-gan}, and the LDPC classic baseline in terms of BLER with the block length of 128 and 256 information bits under the Rician and Rayleigh fading channels, as shown in Fig.~\ref{fig:Rayleigh128} and Fig.~\ref{fig:Rayleigh256}, respectively. As can be observed, the proposed solution shows performance improvement in BLER compared to the baselines under both fading channels. The LDPC baseline performs better than RA-GAN and GAN schemes at low SNR, achieving lower BLER due to its optimization for noisy environments. However, RL-based schemes like AT and DDPG excel at high SNR levels, where they outperform LDPC. This is because DL-based schemes are trained at high SNR values, which may reduce their effectiveness at low SNR but enhance their performance at high SNR. LDPC codes, while effective in low SNR scenarios, become less efficient at high SNR due to increased complexity and decoding latency. As can be observed in Figs. \ref{fig:Rayleigh8}, \ref{fig:Rayleigh128}, and \ref{fig:Rayleigh256}, the proposed DDPG solution can achieve a performance close to that of the supervised learning-based solution without requiring channel knowledge in advance.

The conditional GAN method utilizes conditional GANs to represent the channel effect in wireless communications, thereby connecting the transmitter and the receiver, which allows the gradient to be back-propagated from the receiver NN to the transmitter NN. However, the conditional GAN method cannot guarantee the availability of gradient when training the data with the large dimensional block length. The RA-GAN solution demonstrates slight performance improvement by utilizing ResNet and $\bm{l}_2$ regularization, but it cannot effectively mitigate the local optimization arising from BP. During the BP process from the receiver to the transmitter, via the generator-based channel, there is a certain degree of information loss in the error gradient, potentially leading to performance degradation. In contrast, the RL-based schemes including AT scheme and the proposed DDPG-based approach circumvent the blocked gradient problem by using the loss feedback mechanism to jointly train the transmitter and the receiver networks. Specifically, the DDPG scheme adopts an actor-critic architecture to enable more effective training of the transmitter by leveraging the Q-value estimations generated by the critic network. The critic network is optimized by minimizing the difference between the reward batches and the predicted Q-values, where the reward batch accurately preserves the loss information from the receiver network. This mechanism explains the overall performance improvement of the proposed DDPG solution. Additionally, the experience replay buffer enhances learning stability and improves convergence by enabling the sampling of past experiences, with a focus on those associated with higher reward values.

\subsubsection{Unknown Channel}
Finally, we evaluate the performance of the proposed scheme under 3GPP channel setting \cite{salo2005matlab}. In addition, we consider the case when the channel knowledge is not available in advance, which is common in practice where the channel between the transmitter and the receiver is typically complicated and may not be feasible to formulate. As shown in Fig. \ref{fig:3gpp}, the proposed DDPG-based solution demonstrates performance comparable to the classical AE baseline trained with a known channel. Notably, the performance advantage becomes obvious when compared to the AE baseline trained on an unknown channel. In specific, while the channel layer of the AE is fixed to a Rayleigh fading channel, the actual encoded signal passes through the 3GPP channel. The end-to-end loss at the receiver is used to update the entire AE via BP. This setup reflects the real-world scenario where the channel layer remains complex to be formulated. The results show that both the RL-based schemes and the LDPC scheme outperform AE trained on an unknown channel. In the low SNR range, the LDPC baseline shows the strongest resilience to noise, outperforming the other methods. However, as the SNR increases, both RL-based solutions surpass the LDPC scheme, demonstrating the effectiveness of the proposed CNN architecture and reinforcement learning procedure. The results underscore the ability of the proposed DDPG-based solution to adapt and excel under dynamic and complex channel scenarios.

\section{Conclusion}
\label{sec:conclusion}
In this article, we have proposed a DDPG-based E2E learning solution to relax the requirement of the prior channel model in conventional E2E communication systems. In particular, the proposed CNN architecture can overcome the curse of dimensionality problem in E2E communication systems, which enables data transmission with large block lengths (i.e., 256 bits). With the DDPG algorithm, the transmitter updates its NN by learning from the reward (i.e., the loss value) sent by the receiver, based on the current state (i.e., bitstream) and the chosen action (i.e., encoded symbol). The use of a critic network for Q-value estimation, along with the experience replay buffer, enables the transmitter's NN to more effectively learn the optimal mapping of encoded symbols. This process allows the transmitter to implicitly learn information about the receiver's training, thereby adapting its NN and improving overall system performance. Simulation results across various wireless communication scenarios demonstrate the effectiveness of our proposed solution compared to existing methods. Furthermore, convergence analysis shows that the DDPG-based scheme achieves better convergence than baseline approaches throughout the entire training process.

\bibliographystyle{IEEEtran}
\bibliography{reference.bib}

\end{document}